\documentclass[twocolumn,superscriptaddress,prl,aps]{revtex4-1}
\usepackage{amssymb}
\usepackage{amsmath}
\usepackage{graphicx}
\usepackage{epsfig}

\begin{document}

\title{Bulk-Boundary Correspondence in a Non-Hermitian System in One Dimension with Chiral-Inversion Symmetry}
\author{L. Jin}
\email{jinliang@nankai.edu.cn}
\author{Z. Song}
\affiliation{School of Physics, Nankai University, Tianjin 300071, China}

\begin{abstract}
Asymmetric coupling amplitudes effectively create an imaginary gauge field,
which induces a non-Hermitian Aharonov-Bohm (AB) effect. Nonzero imaginary
magnetic flux invalidates the bulk-boundary correspondence and leads to a
topological phase transition. However, the way of non-Hermiticity appearance
may alter the system topology. By alternatively introducing the
non-Hermiticity under symmetry to prevent nonzero imaginary magnetic flux,
the bulk-boundary correspondence recovers and every bulk state becomes
extended; the bulk topology of Bloch Hamiltonian predicts the (non)existence
of edge states and topological phase transition. These are elucidated in a
non-Hermitian Su-Schrieffer-Heeger model, where chiral-inversion symmetry
ensures the vanishing of imaginary magnetic flux. The average value of Pauli
matrices under the eigenstate of chiral-inversion symmetric Bloch
Hamiltonian defines a vector field, the vorticity of topological defects in
the vector field is a topological invariant. Our findings are applicable in other non-Hermitian systems. We first uncover
the roles played by non-Hermitian AB effect and chiral-inversion symmetry
for the breakdown and recovery of bulk-boundary correspondence, and develop new
insights for understanding non-Hermitian topological phases of matter.
\end{abstract}

\maketitle

\textit{Introduction.---}Topological theory has been well established in
condensed matter physics \cite%
{Kitaev,RyuPRL2002,Greiner,SCZhang,Bernevig06,Kane,LFuPRB,LFuPRL,Ludwig,HJZhang09,RyuNJP,KaneRMP,XLQi,Xu,Burkov,Young,Wang,Wang1,Bardyn2012,Esslinger,Weng,LLuScience,XJLiu,DLD,Liu,CTChanNP,ZLiu13,Leykam2,Ryu,Kunst2017,Kunst2018,Armitage}
and recent experimental progresses in optics boost the development of
topological photonics~{\cite%
{Hafezi2011,BlochPRL,BlochNP,LLuRev,Hafezi2016,LinNC2016,Goldman,Ozawa,Cooper}%
. The existence of gapless edge states of a system under open boundary
condition (OBC) is predictable from the change of topological invariants
associated with the bulk topology of system under periodical boundary
condition (PBC), known as the (conventional) bulk-boundary correspondence,
which is ubiquitously applicable in Hermitian systems. }

In parallel, non-Hermitian physics exhibits considerable intriguing features
\cite%
{BenderRPP,Ali,Makris08,Klaiman08,YDChong10,Moiseyev,Liertzer,Bender,Rotter,UP2013,Cao,LFeng14,Hodaei14,Monticone,Musslimani,AGuo,Ruter,Peng,LChang,Fleury,CTChanPRX,SHFanEPRing,Zhang,EPSensing2,EPSensing3,LGePRA2017,Ashida,SFanNature,KawabataPRL,ZhenScience,Harari,Bandres,PTRev2017,PTRev2018}%
; the unexpected novel interface states appear between non-Hermitian
periodic media with distinct topologies \cite%
{Zoller2011,YCHu,Esaki,GQLiang,Schomerus,Chen,Rudner,Zeuner,Yuce,HZhao,WangX,Poli,CCSymmetry,Weimann}%
. These stimulate the studies of topological phases and edge states in
non-Hermitian systems \cite%
{LFengLaser,LFengNC2018,Parto,PXue,LJin,Downing,Klett,HMenke,LJinPRA2017,Leykam,YXu,SFan2018,Lieu,Yuce2018,ChongPRB,MolinaPRB,SChen2018PRA,LFu,LiangFUPRL2018,Kawabata,WR,KlettArxiv,LieuArxiv,KawabataArxiv,HJiangArxiv,MalzardArxiv,HChen,LJLArxiv}%
. Non-Hermitian band theory and the topological characterization are
developed employing the left and right eigenstates \cite{Rotter,Moiseyev};
the Chern number, generalized Berry phase and winding numbers are quantized
as topological invariants \cite{Leykam,YXu,LFu}.

Remarkably, the bulk-boundary correspondence \cite{BBC} is invalid in
certain non-Hermitian topological systems \cite{TonyLeePRL,Xiong2018,Torres}%
: Systems under PBC and OBC have dramatically different energy spectra, and
all the eigenstates localize near system boundaries (the non-Hermitian skin
effect) \cite{ZhongWang,Kunst,ZGongPRX}. These have received great research
interests in non-Hermitian systems of asymmetric Su-Schrieffer-Heeger (SSH)
model, topological insulators, and nodal-line semimetals \cite%
{Kunst,ZhongWang,YaoArxiv,ZGongPRX,CHLee,HWang,ZYang,TorresReview,KawabataCI,SSH}%
. Biorthogonal \cite{Kunst} and non-Bloch bulk-boundary correspondences \cite%
{ZhongWang} are suggested. In contrast, non-Hermiticity does not inevitably
destroy the bulk-boundary correspondence \cite%
{YCHu,Esaki,Lieu,LJLArxiv,Kawabata,KawabataArxiv,WR}, which is verified in a
parity-time-symmetric non-Hermitian SSH model with staggered couplings and
losses \cite%
{Rudner,Schomerus,Zeuner,Poli,Weimann,LFengLaser,LFengNC2018,Parto}. {%
Questions arise: Why bulk-boundary correspondence fails in certain
non-Hermitian systems? What roles do non-Hermiticity and symmetry play in
the breakdown of bulk-boundary correspondence? How to characterize the
topological properties and understand the topological invariant without
(conventional) bulk-boundary correspondence? }

In this Rapid Communication, we first report that \textit{chiral-inversion
symmetry plays an important role} for the bulk-boundary correspondence in
non-Hermitian system of a non-Hermitian SSH model with asymmetric coupling;
which leads to a \textit{non-Hermitian Aharonov-Bohm (AB) effect} with an
imaginary magnetic flux under PBC and a non-Hermitian skin effect under OBC
without chiral-inversion symmetry. The imaginary magnetic flux results in
complex spectrum and a topological phase transition, but it disappears under
OBC; the OBC spectrum significantly differs from the PBC spectrum, and the
bulk-boundary correspondence fails. Non-Hermitian AB effect vanishes if the
asymmetry is alternatively introduced without breaking chiral-inversion
symmetry. The bulk-boundary correspondence is valid; a topological invariant
is constructed from the system bulk with the imaginary gauge field removed,
being the vorticity of band touching points as topological defects in the
vector field defined from the average values of Pauli matrices. Our findings are valid for other non-Hermitian
topological systems.

\begin{figure}[b]
\includegraphics[ bb=0 0 273 230, width=0.46\textwidth, clip]{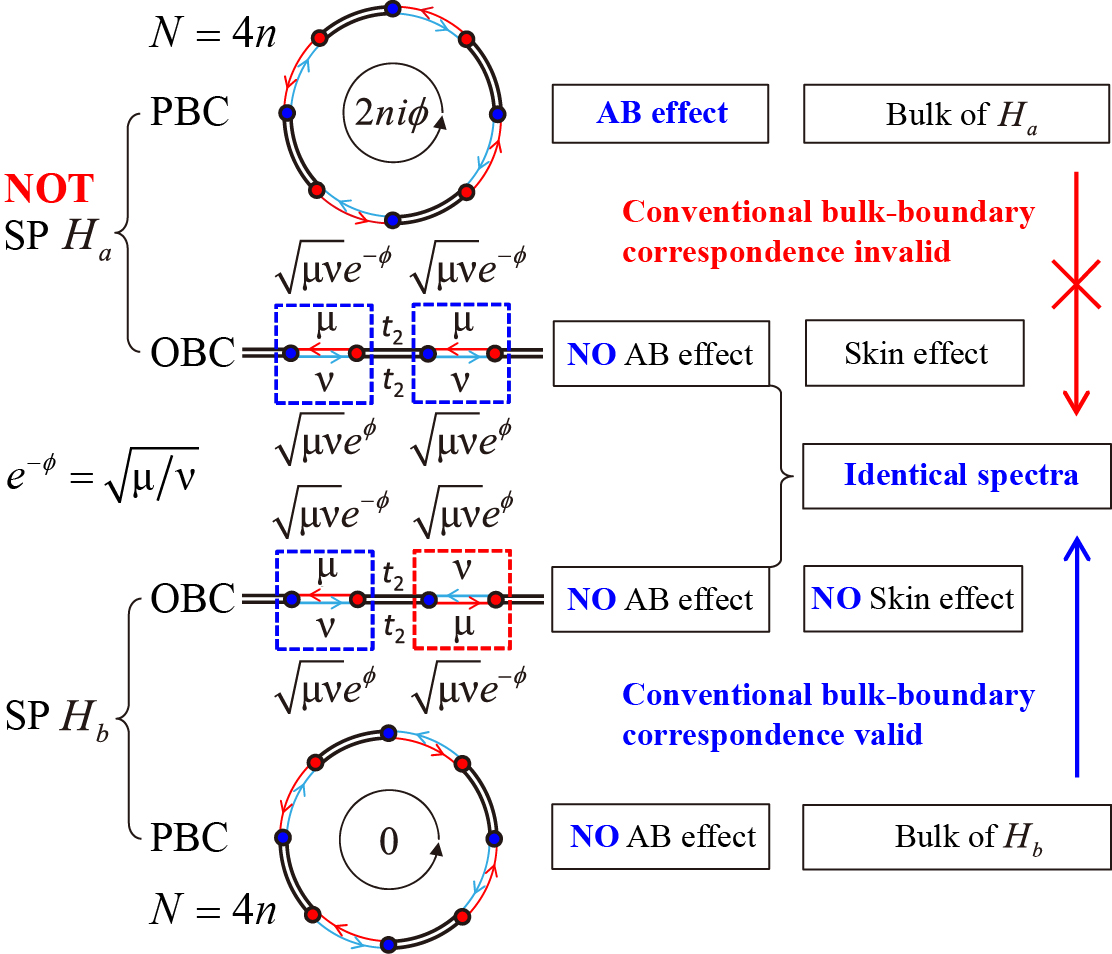}
\caption{Breakdown (recovery) of bulk-boundary correspondence for system without (with) chiral-inversion symmetry
from the viewpoint of non-Hermitian AB effect. Imaginary gauge field induces nonzero (zero) imaginary magnetic flux in $H_a$ ($H_b$) under PBC. Topological invariant obtained from the bulk Bloch Hamiltonian of system $b $ is a Bloch (non-Bloch) topological invariant for system $b$ ($a$).
Lattice size is $N=4n$.}
\label{fig1}
\end{figure}

\textit{Topological phase transition induced by symmetry breaking.}---The
Bloch Hamiltonian of a non-Hermitian system $a$ (Fig. \ref{fig1}) under PBC
is
\begin{equation}
H_{a}\left( k\right) =\left( t_{1}+t_{2}\cos k\right) \sigma _{x}+\left(
t_{2}\sin k-i\gamma \right) \sigma _{y},  \label{1}
\end{equation}%
where $\sigma _{x,y}$ are the Pauli matrices. $t_{2}$ is the intercell
coupling. Set $\mu =t_{1}-\gamma $ and $\nu =t_{1}+\gamma $, the asymmetric
intracell coupling amplitude ($\mu \neq \nu ^{\ast }$) raises the
non-Hermiticity. Non-Hermitian asymmetric coupling can be realized between
primary resonators evanescently coupled through auxiliary resonator~\cite%
{Hatano,SLonghiExp,MidyaFengNC}, which has half perimeter gain and half
perimeter loss, leading to the amplification and attenuation for the
coupling amplitudes in opposite tunneling directions. Implementation of
asymmetric coupling with ultracold atoms in optical lattice is possible~\cite%
{ZGongPRX}.

In Hermitian case ($\gamma =0$), system $a$ holds chiral-inversion symmetry%
\begin{equation}
\left( \mathcal{SP}\right) H_{a}\left( k\right) \left( \mathcal{SP}\right)
^{-1}=-H_{a}\left( -k\right) ,U_{\mathcal{SP}}H_{a}U_{\mathcal{SP}%
}^{-1}=-H_{a}.  \label{2}
\end{equation}%
The constraints are for a \textit{combined} chiral-inversion symmetry. $%
\mathcal{SP}$ and $U_{\mathcal{SP}}$ are unitary operators. $H_{a}$ [$%
H_{a}(k)$] is the Hamiltonian in the real-space ($k$-space). Two band
touching degeneracy points exist [Fig~\ref{fig2}(d)].

In non-Hermitian case ($\gamma \neq 0$), unlike the alternative gain and
loss \cite{Schomerus,Zeuner,Poli,Weimann,LFengLaser,LFengNC2018,Parto}, the
asymmetric coupling breaks the chiral-inversion symmetry ($H_{a}$ in Fig.~%
\ref{fig1}). Taken $\mu \nu >0$ as illustration (see
Supplemental Material A \cite{SI}) and rewritten
\begin{equation}
\mu =\sqrt{\mu \nu }e^{-\phi },\nu =\sqrt{\mu \nu }e^{\phi },  \label{munu}
\end{equation}
where $e^{-\phi }\equiv \sqrt{\mu /\nu }$ \cite%
{XZZhang,SLonghiExp,MidyaFengNC}, the asymmetric coupling is expressed as a
symmetric coupling $\sqrt{\mu \nu }$ with Peierls \textquotedblleft phase"
factor~\cite%
{Hafezi2011,BlochPRL,Goldman,Ozawa,Cooper,KFang,KFangNPhoton,ELi,HafeziPRL2014,LJin16,LJin17,LJin18}
of amplification/attenuation $e^{\pm \phi }$~\cite{RemarkPeriels}, which
indicates the presence of an imaginary gauge field \cite%
{Hatano,SLonghiExp,MidyaFengNC}. A non-Hermitian AB \textquotedblleft phase"
factor of amplification/attenuation $e^{\pm i(2ni\phi )}$ is accumulated
when particle circling a loop in $H_{a}$ under PBC; where $2ni\phi $ is the
imaginary magnetic flux~\cite%
{HafeziPRL2014,LJin16,LJin17,LJin18,RemarkPeriels}. The eigenvalues are%
\begin{equation}
E_{a,\pm }=\pm \sqrt{t_{2}^{2}+\mu \nu +2t_{2}\sqrt{\mu \nu }\cos \left(
k+i\phi \right) },
\end{equation}%
with $k=\pi m/n$, integer $m\in \lbrack 1,2n]$ [Fig.~\ref{fig2}(a)].

\begin{figure}[b]
\includegraphics[ bb=0 0 520 385, width=0.49\textwidth, clip]{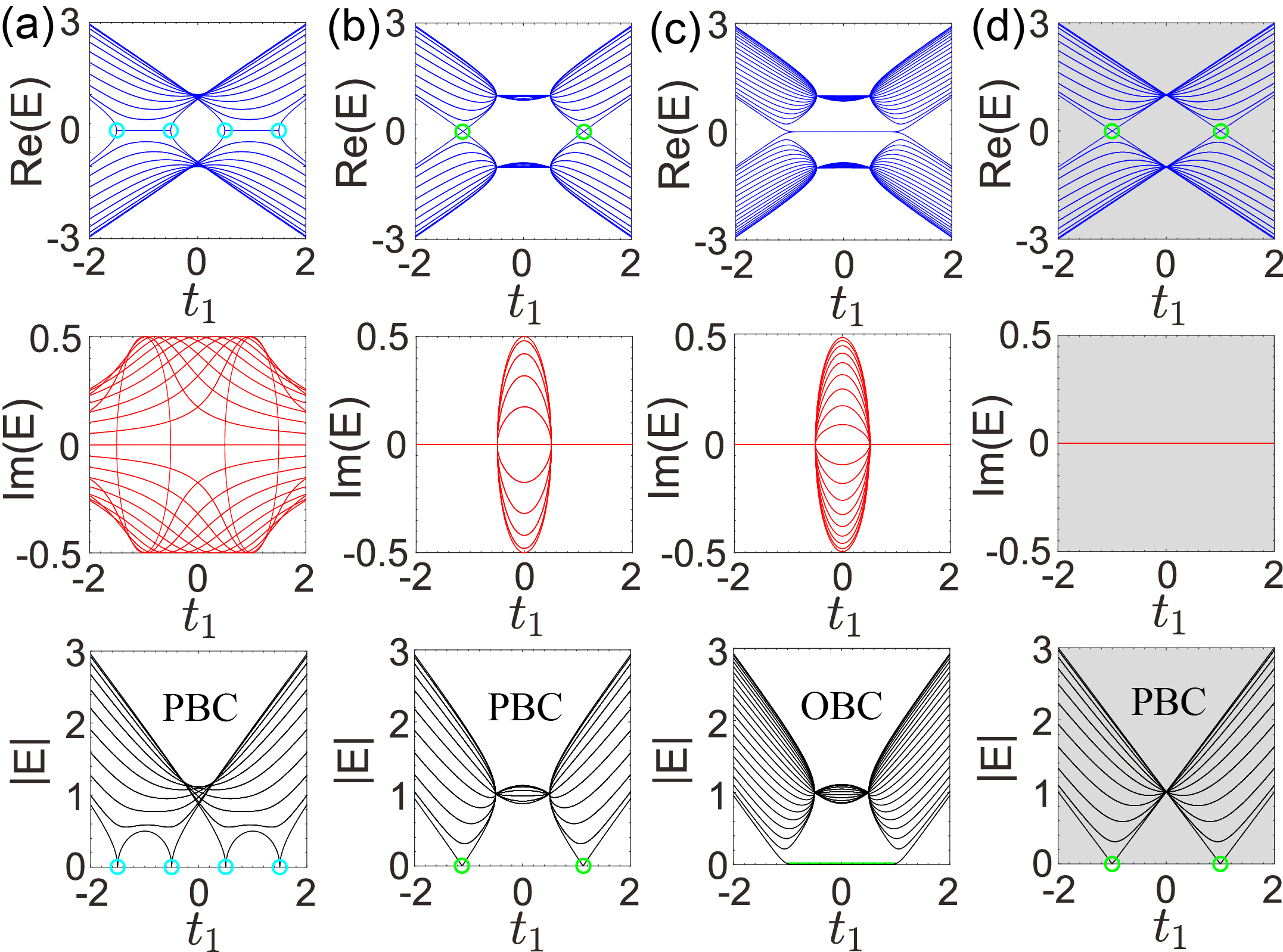}
\caption{(a) [(b)] Energy spectrum for $H_a$ ($H_b$) under PBC. (c) Identical spectra of $H_a$ and $H_b$ under OBC with one intercell coupling $t_2$ missing, and (d) $H_a$ and $H_b$ under PBC in the Hermitian case ($\gamma=0$). The band touching exceptional points (degeneracy points) are indicated by the cyan (green) hollow circles.
The system parameters are $N=40$, $t_2=1$, and $\gamma=1/2$ in (a-c).} \label%
{fig2}
\end{figure}

In contrast to a real magnetic flux that shifts $k$ in the momentum space
without varying the dispersion relation \cite{HafeziPRL2014}, the momentum
changes to $k+i\phi$~\cite{YaoArxiv,ZYang,CHLee} and spectrum becomes fully
complex affected by imaginary magnetic flux; which induces a topological
phase transition with band touching degeneracy points splitted into pairs of
band touching EPs [Figs.~\ref{fig2}(a) and~\ref{fig2}(d)] \cite{ZhenScience}
that exhibit different topology~\cite%
{Heiss2001,DembowskiPRL2001,Berry,GP,Rotter2008,Uzdin,Heiss,Huang2013,BZhen,Doppler,HXu,RiemannSheet,Hassan,Goldzak,CTChanPRX2018,CTChanPRL2018,LJinPRL2018}%
. Imaginary magnetic flux is absence under OBC, thus the spectra and band
touching points under PBC [Fig.~\ref{fig2}(a)] and OBC [Fig.~\ref{fig2}(c)]
are dramatically different \cite%
{Torres,TonyLeePRL,Xiong2018,Kunst,ZhongWang,ZGongPRX,CHLee,HWang,ZYang,TorresReview}%
. The eigenstate amplitude is one-way enlarged under OBC because of
imaginary gauge field~\cite{SLonghiExp,MidyaFengNC,RemarkBiorthgonal}; and
all the eigenstates localize at system boundary (non-Hermitian skin effect~%
\cite{ZhongWang,ZGongPRX,CHLee,ZYang,HWang}). The localization length is $%
\xi =\phi ^{-1}$ \cite{LGeLL}. The inverse participation ratio (IPR) $%
\sum_{j}|\psi _{j}|^{4}/ (\sum_{j}|\psi _{j}|^{2})^2$ of bulk states scales
as $N^{-1}$ for small $N$, particularly for weak non-Hermiticity; and
becomes system size insensitive when the localization dominates at large $N$
(see Supplemental Material B \cite{SI}).

\textit{Bulk-boundary correspondence.}---Chiral-inversion symmetry holds
when non-Hermiticity is alternatively introduced in system $b$ ($H_b$ in
Fig.~\ref{fig1}) \cite{ChiralInversionSymmetry}. Under symmetry protection,
two degeneracy points move without splitting into EP pairs [Figs. \ref{fig2}%
(b) and \ref{fig2}(d)]. The eigenstates under OBC are
symmetric/antisymmetric. All bulk states are extended and non-Hermitian skin
effect disappears even most bulk states have complex eigenvalues (IPR of
bulk states of system $b$ is inversely proportional to the system size \cite%
{SI}). Although the significant difference between eigenstates, systems $a$
and $b $ under OBC possess \textit{identical} energy spectra (see Supplemental Material C \cite{SI}), the imaginary gauge fields do not
affect OBC spectra [Fig.~\ref{fig2}(c)]. These manifest that \textit{the way
of non-Hermiticity appearance affects system topology}. In particular, the%
\textit{\ non-Hermiticity solely induces nontrivial topology} at $%
t_{1}=t_{2} $~\cite{Takata}.

The amplification and attenuation cancel in $H_b$. The \textit{combined}
chiral-inversion ($\mathcal{SP}$) symmetry prevents the appearance of
nonzero imaginary magnetic flux and the bulk-boundary correspondence is
valid [Figs. \ref{fig2}(b) and \ref{fig2}(c)] (also in Refs. \cite%
{YCHu,Esaki,Schomerus,Zeuner,Poli,Weimann,LFengLaser,LFengNC2018,Parto,Lieu,WR}%
, but is invalid in Refs. \cite%
{Torres,TonyLeePRL,Xiong2018,Kunst,ZhongWang,ZGongPRX,CHLee,HWang,ZYang,TorresReview,YaoArxiv,KawabataCI}
without chiral-inversion symmetry), while \textit{individual} chiral and
inversion symmetries are not necessarily hold separately (see
Supplemental Material D \cite{SI}). The (non)existence of topologically
protected edge states is predictable from the bulk of system $b$,
\begin{equation}
H_{b}\left( k\right) =\left(
\begin{array}{cccc}
0 & \sqrt{\mu \nu }e^{-\phi } & 0 & t_{2}e^{-ik} \\
\sqrt{\mu \nu }e^{\phi } & 0 & t_{2} & 0 \\
0 & t_{2} & 0 & \sqrt{\mu \nu }e^{\phi } \\
t_{2}e^{ik} & 0 & \sqrt{\mu \nu }e^{-\phi } & 0%
\end{array}%
\right) .  \label{Hb}
\end{equation}%
Through a similar transformation with only nonzero diagonal elements $U_{\mu
\nu }=\mathrm{diag}\left( \sqrt{\nu },\sqrt{\mu },\sqrt{\mu },\sqrt{\nu }%
\right) $, the imaginary gauge fields (factors $e^{\pm \phi }$) are removed
from $H_{b}$ (see Supplemental Material E \cite{SI}); and we
obtain $U_{\mu \nu }H_{b}\left( k\right) U_{\mu \nu }^{-1}$, which is
equivalent to a two-site unit cell bulk $h_{b}\left( k\right) =\left( \sqrt{%
\mu \nu }+t_{2}\cos k\right) \sigma _{x}+(t_{2}\sin k)\sigma _{y}$. The
eigenvalues are
\begin{equation}
E_{b,\pm }=\pm \sqrt{t_{2}^{2}+\mu \nu +2t_{2}\sqrt{\mu \nu }\cos \left(
k\right) },  \label{Eb}
\end{equation}
where $k=\pi m/n$, $m\in \lbrack 1,2n]$. The bulk topology of $h_{b}\left(
k\right) $ correctly predicts the (non)existence of edge states in both
systems $a$ and $b$ under OBC (Fig.~\ref{fig1}) \cite{SI}. Removing the
imaginary gauge field in system bulk gives $h_{b}\left( k\right) $, which is
identical with that found by solving the open system~\cite{ZhongWang}.

For $\gamma =|r|e^{i\theta }$ $\left( -\pi \leq \theta \leq \pi \right) $,
the band gap closes at
\begin{equation}
(t_{1}^{2}-|r|^{2})^{2}+4t_{1}^{2}|r|^{2}\sin ^{2}\theta =t_{2}^{4},
\end{equation}%
and $\cos ^{2}\left( k\right) =[t_{2}^{2}+t_{1}^{2}-|r|^{2}\cos \left(
2\theta \right) ]/(2t_{2}^{2})$. The finite size effects appear in discrete
systems (see Supplemental Material F \cite{SI}). For real $\mu
$ and $\nu $ at $\theta =0$, the band touching points are degeneracy
(exceptional) points at $t_{1}^{2}=+\left( -\right) t_{2}^{2}+\gamma ^{2}$
\cite{Kunst,ZhongWang}, being topological defects carrying integer
(half-integer) vorticity. The band touching EPs only appear for $\gamma
^{2}>t_{2}^{2}$.

\begin{figure}[b]
\includegraphics[ bb=0 0 365 350, width=0.49\textwidth, clip]{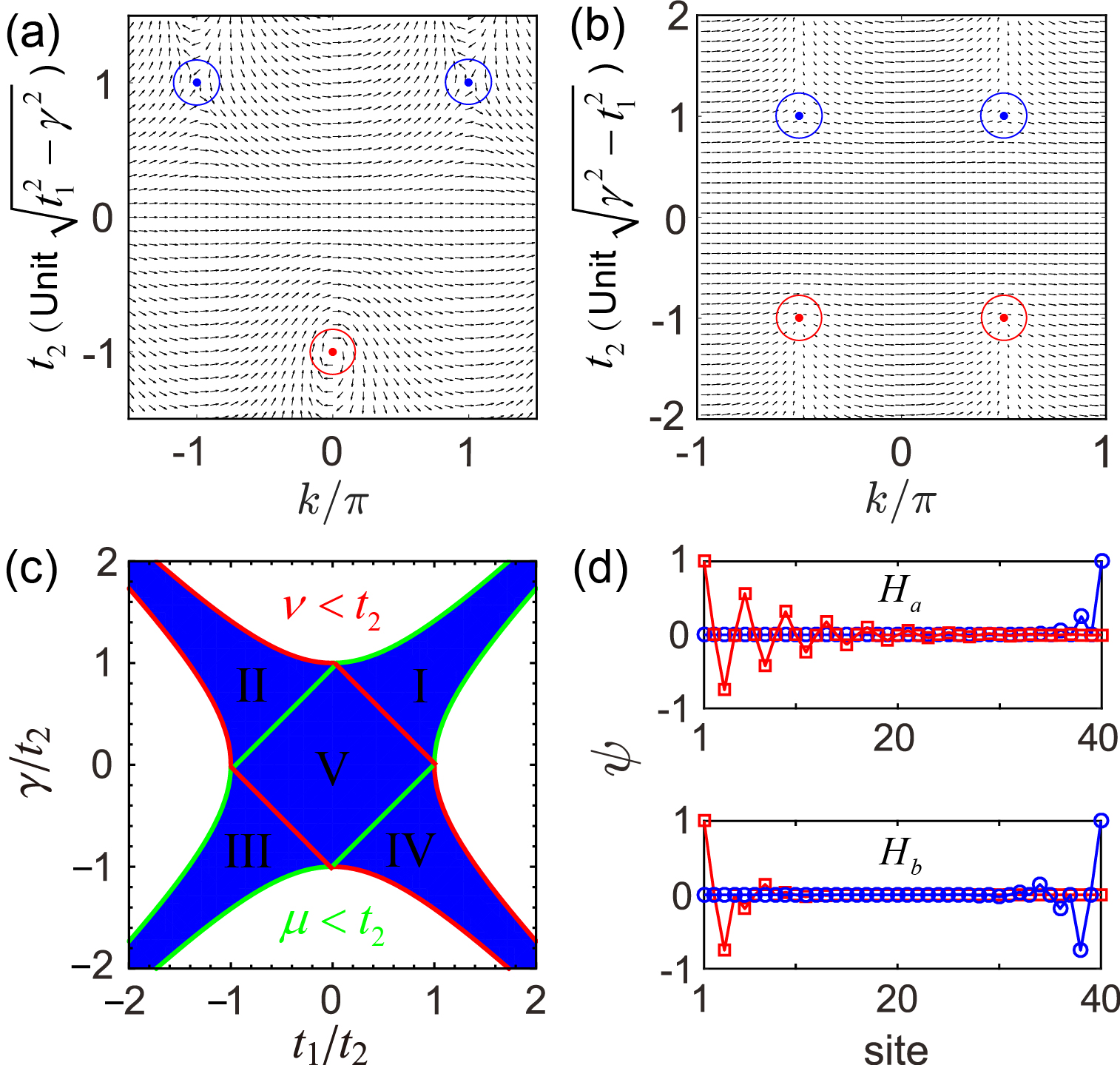}
\caption{Vector field $\mathbf{F}(\mathbf{k})=\left(
\left\langle \sigma _{x}\right\rangle ,\left\langle \sigma _{y}\right\rangle
\right) $ associated with $E_{b,+}$ of $h_b(k)$ for (a) $\mu \nu=t_1^2-\gamma^2>0$ and (b) $\mu \nu=t_1^2-\gamma^2<0$. Red (blue) circles indicate the topological defects with vortices (antivortices), which appear at $( k,t_{2}) =( 0,-\sqrt{t_1^2-\gamma^2
}) $ or $( \pm \pi ,\sqrt{t_1^2-\gamma^2}) $ in (a) and at $( k,t_{2}) =( \pm \pi /2,\pm
\sqrt{\gamma^2-t_1^2}) $ in (b).
(c) Phase diagram for real $\gamma$. Two topological zero edge states exist in the blue region $-t_2^2<\mu \nu< t_2^2$ for one intercell coupling $t_2$ missing.
(d) Zero edge states for systems $a$ and $b$ under OBC. The system parameters are $N=40$, $t_1=1/4$,
$\gamma=1/2$, and $t_2=1$.} \label{fig3}
\end{figure}

\textit{Topological invariant.---}Topology invariants are recently
constructed in non-Hermitian systems \cite%
{Leykam,LFu,YXu,ZhongWang,SChen2018PRA,WR}. The Chern number defined via
Berry curvature \cite{YXu,WR}, the vorticity defined via the complex energy
\cite{LFu}, and the generalized Berry phase defined via the argument of
effective magnetic field \cite{Leykam,YXu,SChen2018PRA} are quantized. The
vorticity of topological defects in a vector field $\mathbf{B}(\mathbf{k})$
associated with the Bloch Hamiltonian is a topological invariant \cite%
{Sarma,Hou}; we generalize this vorticity to non-Hermitian systems through
defining a two-component vector field $\mathbf{F}(\mathbf{k})=\left(
\left\langle \sigma _{x}\right\rangle ,\left\langle \sigma _{y}\right\rangle
\right) $ [Figs.~\ref{fig3}(a) and~\ref{fig3}(b)] that is composed by the
average values of Pauli matrices under the eigenstates of $h_{b}\left(
k\right) $. $w=\oint_{L}\left( 2\pi \right) ^{-1}(\hat{F}_{x}\nabla \hat{F}%
_{y}-\hat{F}_{y}\nabla \hat{F}_{x})\mathrm{d}\mathbf{k}$ characterizes the
vorticity of topological defects inside the loop $L$ in the parameter plane $%
\mathbf{k=}\left( k,t_{2}\right) $, where $\hat{F}_{x(y)}=F_{x(y)}/\sqrt{%
F_{x}^{2}+F_{y}^{2}}$ and $\nabla =\partial /\partial \mathbf{k}$, which is
in accord with that defined in the Brillouin zone of a two-dimensional (2D)
brick wall lattice (see Supplemental Material E \cite{SI}).
The varying direction of $\mathbf{F}(\mathbf{k})$ accumulated is $2\pi w=\pm
2\pi $ ($\pm \pi $) in Fig.~\ref{fig3}(a) [Fig.~\ref{fig3}(b)] if $L$
encircles a topological defect, the plus (minus) sign corresponds to the
vortex (antivortex); otherwise, if $L$ does not encircle a topological
defect, the varying direction is $2\pi w=0$.

Phase diagram is plotted in Fig.~\ref{fig3}(c) for real $\gamma $. For $\mu
\nu >0$, the degeneracy points are at $t_{2}^{2}-\mu \nu =0$. As
non-Hermiticity increases, the band gap inside two EPs \cite%
{Rotter,PTRev2017,PTRev2018} with complex spectrum diminishes and closes at $%
t_{1}=0$ when $\gamma ^{2}=t_{2}^{2}$. $\mu \nu =0$ ($t_{1}=\pm \gamma $)
are EPs, where the eigenstates are highly defective and fully constituted by
two-state coalescences at energy $\pm t_{2}$. For $\mu \nu <0$, $%
t_{2}^{2}+\mu \nu =0$ yields another boundary for the zero edge states
determined from band touching EPs. Two topological zero edge states exist in
the regions $\gamma ^{2}-t_{2}^{2}<t_{1}^{2}<\gamma ^{2}+t_{2}^{2}$ for one
intercell coupling $t_{2}$ vanishing under OBC \cite{NoteBoundary}.

\textit{Topological edge states.}---The bulk topology relates to the
(dis)appearance of edge states at the interfaces where topological invariant
($w$) changes. We consider systems with complete unit cells ($N=4n$) with
one $t_{2}$ vanishes (see Supplemental Material G for the case with a defective unit cell~\cite{SI}). In system $b$, two edge states localize on
the left and right boundaries, respectively in all blue regions of Fig.~\ref%
{fig3}(c). In system $a$, this occurs only in region $\mathrm{V}$; and both
two edge states localize on the right (left) boundary in regions $\mathrm{I}$
and $\mathrm{III}$ ($\mathrm{II}$ and $\mathrm{IV}$).

For system $b$, the left edge state is $\psi _{2j}=0$ and
\begin{equation}
\psi _{2j+1}=-[\left( \mu +\nu \right) +\left( -1\right) ^{j}\left( \mu -\nu
\right) ]/\left( 2t_{2}\right) \psi _{2j-1},
\end{equation}%
at $N\gg 1$. The right edge state is a left-right spatial reflection of the
left edge state [Fig.~\ref{fig3}(d)]. Anomalous edge states localize in one
unit cell at system boundary at the EPs ($t_{1}^{2}=\gamma ^{2}$)~\cite%
{TonyLeePRL,Torres,Leykam}. At $t_{1}=-\gamma $, the left (right) edge state
is $\psi _{1}=1 $ ($\psi _{N}=1 $); at $t_{1}=\gamma $, the left edge state
is $\psi _{1}=-\left( +\right) \psi _{3}=1$ and the right edge state is $%
\psi _{N}=-\left( +\right) \psi _{N-2}=1$ when $t_{1}/t_{2}>0$ ($%
t_{1}/t_{2}<0$).

In contrast, for system $a$, the left edge state is $\psi _{2j}=0$ and $\psi
_{2j+1}=\left( -\nu /t_{2}\right) \psi _{2j-1}$; the right edge state is $%
\psi _{2j-1}=0$ and $\psi _{N-2j}=\left( -\mu /t_{2}\right) \psi _{N+2-2j}$
with a different decay rate $-\mu /t_{2}$ [Fig.~\ref{fig3}(d)]. The
imaginary gauge field induces imbalanced probability distributions between
edge states. The green (red) ribbon in Fig.~\ref{fig3}(c) indicates $%
\left\vert \mu /t_{2}\right\vert <1$ ($\left\vert \nu /t_{2}\right\vert <1$%
), both edge states localize on the right (left) boundary. The edge states
are $\psi _{1}=1$ ($\psi _{N}=1$) for $t_{1}/t_{2}<0$ ($t_{1}/t_{2}>0$) at
the EPs.

\begin{figure}[tb]
\includegraphics[ bb=0 0 535 160, width=8.7 cm, clip]{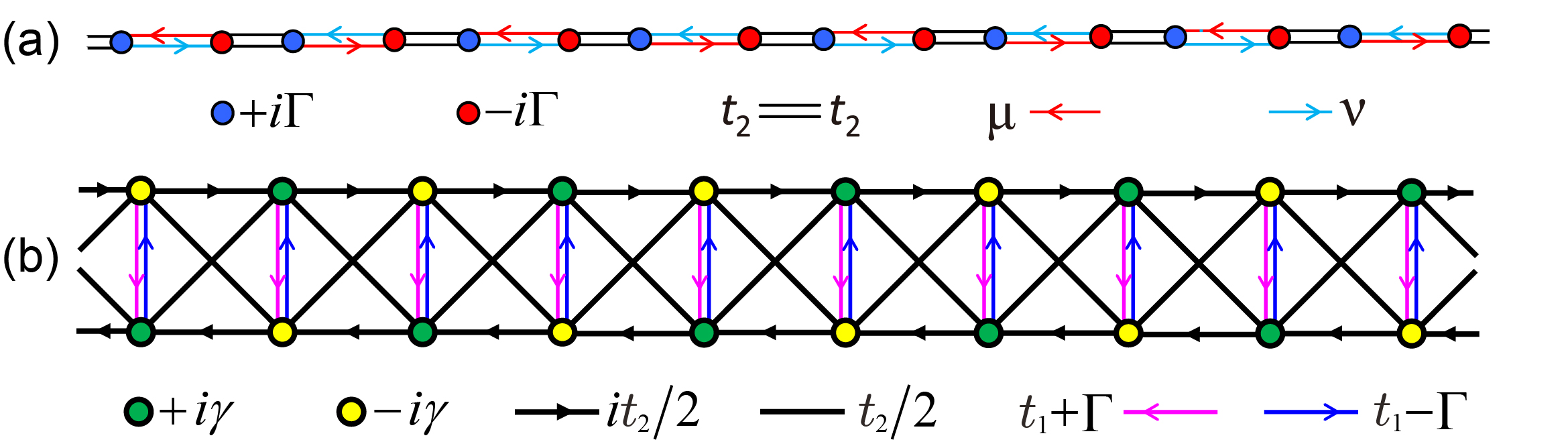}
\caption{ Chiral-inversion symmetric (a) non-Hermitian SSH model of system $b$ in Fig.~\ref{fig1} ($H_b$)
with staggered gain and loss $\Gamma$, (b) non-Hermitian Creutz ladder that equivalent to (a).} \label{fig4}
\end{figure}

\textit{Discussion and Conclusion.}---{Figure~\ref{fig4}(a)
depicts the chiral-inversion symmetric non-Hermitian SSH model of system $b$ with staggered gain and loss $\Gamma$, where chiral
symmetry and inversion symmetry are not separately hold. System shown in Fig.~%
\ref{fig4}(a) is equivalent to the chiral-inversion symmetric non-Hermitian Creutz ladder [Fig.~\ref%
{fig4}(b)]. The Creutz ladder has} a $\pi $ magnetic flux in each plaquette
\cite{Creutz}. The Creutz ladder in Refs. \cite{TonyLeePRL,Xiong2018,Torres} is
equivalent to system $a$ through a similar transformation $U=I_{2n}\otimes
\left( i\sigma _{x}+I_{2}\right) $ (see
Supplemental Material H \cite{SI}), where bulk-boundary
correspondence fails because gain and loss associated with real magnetic
flux breaks chiral-inversion symmetry and effectively creates imaginary
magnetic flux under PBC.

Time-reversal (Inversion) symmetry prevents nonzero real (imaginary)
magnetic flux. An attenuation (amplification) factor $e^{-\phi }$
accompanied with the corresponding amplification (attenuation) factor $%
e^{\phi } $ in the direction concerned can prevent nonzero imaginary
magnetic flux. This is enabled under inversion symmetry ($\mathcal{P}$) or
combined inversion symmetries such as chiral-inversion ($\mathcal{SP}$)
symmetry, charge-conjugation inversion ($\mathcal{CP}$) symmetry, and
parity-time ($\mathcal{PT}$) symmetry. For a 2D non-Hermitian Chern
insulator $\left( m+t\cos k_{x}+t\cos k_{y}\right) \sigma _{x}+\left( t\sin
k_{x}+i\gamma \right) \sigma _{y}+t\sin k_{y}\sigma _{z}$~\cite{KawabataCI},
we write its energy bands as $\pm \sqrt{\mu \nu +t^{2}+t^{2}\sin
^{2}k_{y}+2t\sqrt{\mu \nu }\cos \left( k_{x}+i\phi \right) }$, where we set $\mu =m+t\cos k_{y}+\gamma $, $\nu =m+t\cos k_{y}-\gamma $, and $%
\sqrt{\mu /\nu }=e^{-\phi }$ (for $\mu \nu >0$). An imaginary
magnetic flux exists in the $x$ direction, but not in the $y$ direction;
considerable difference between PBC and OBC spectra is observed in the $x$
direction~\cite{KawabataCI}. Introducing the non-Hermiticity under inversion
symmetry prevents nonzero imaginary magnetic flux under PBC because of
the cancellation between amplification and attenuation factors $e^{\pm \phi }
$ in the $x$ direction and enables the bulk-boundary correspondence. By applying the same procedure done for
the non-Hermitian SSH model of system $b$, we can obtain an equivalent bulk
Bloch Hamiltonian $h_{b,CI}\left( k_{x},k_{y}\right) =( \sqrt{\mu \nu }%
+t\cos k_{x} ) \sigma _{x}+(t\sin k_{x})\sigma _{y}+(t\sin k_{y})\sigma _{z}$
after removing the imaginary gauge field (see Supplemental Material I \cite{SI}). The
energy bands are $\pm \sqrt{\mu \nu +t^{2}+t^{2}\sin ^{2}k_{y}+2t\sqrt{\mu
\nu }\cos k_{x}}$. The bulk topology of $%
h_{b,CI}\left( k_{x},k_{y}\right) $ correctly predicts the topological
phase transition and the (non)existence of edge states for the Chern
insulator under OBC.

Bulk-boundary correspondence fails for nonzero imaginary magnetic flux under
PBC if the flux vanishes under OBC; the bulk-boundary correspondence
recovers by alternatively introducing non-Hermiticity under symmetry, which
prevents nonzero imaginary magnetic flux; and topological invariant can be
constructed from the bulk Bloch Hamiltonian. The non-Bloch topological
invariant and exotic bulk-boundary correspondence~\cite{ZhongWang,ZGongPRX}
are elaborated from the viewpoint of (conventional) bulk-boundary
correspondence. Our findings provide new insights from non-Hermitian AB
effect and shed light on non-Hermitian topological phases of matter.

\acknowledgments We acknowledge the support of NSFC (Grant Nos. 11605094 and
11874225).

\clearpage
\begin{widetext}

\section*{Supplemental Material for ``Bulk-Boundary Correspondence in a
Non-Hermitian System in One Dimension with Chiral-Inversion Symmetry"}

\begin{center}
L. Jin and Z. Song\\[2pt]
\textit{School of Physics, Nankai University, Tianjin 300071, China}
\end{center}

\subsection*{A. Imaginary gauge field and non-Hermitian Aharonov-Bohm effect}

\begin{figure}[tbh]
\includegraphics[ bb=0 0 535 275, width=0.5\textwidth, clip]{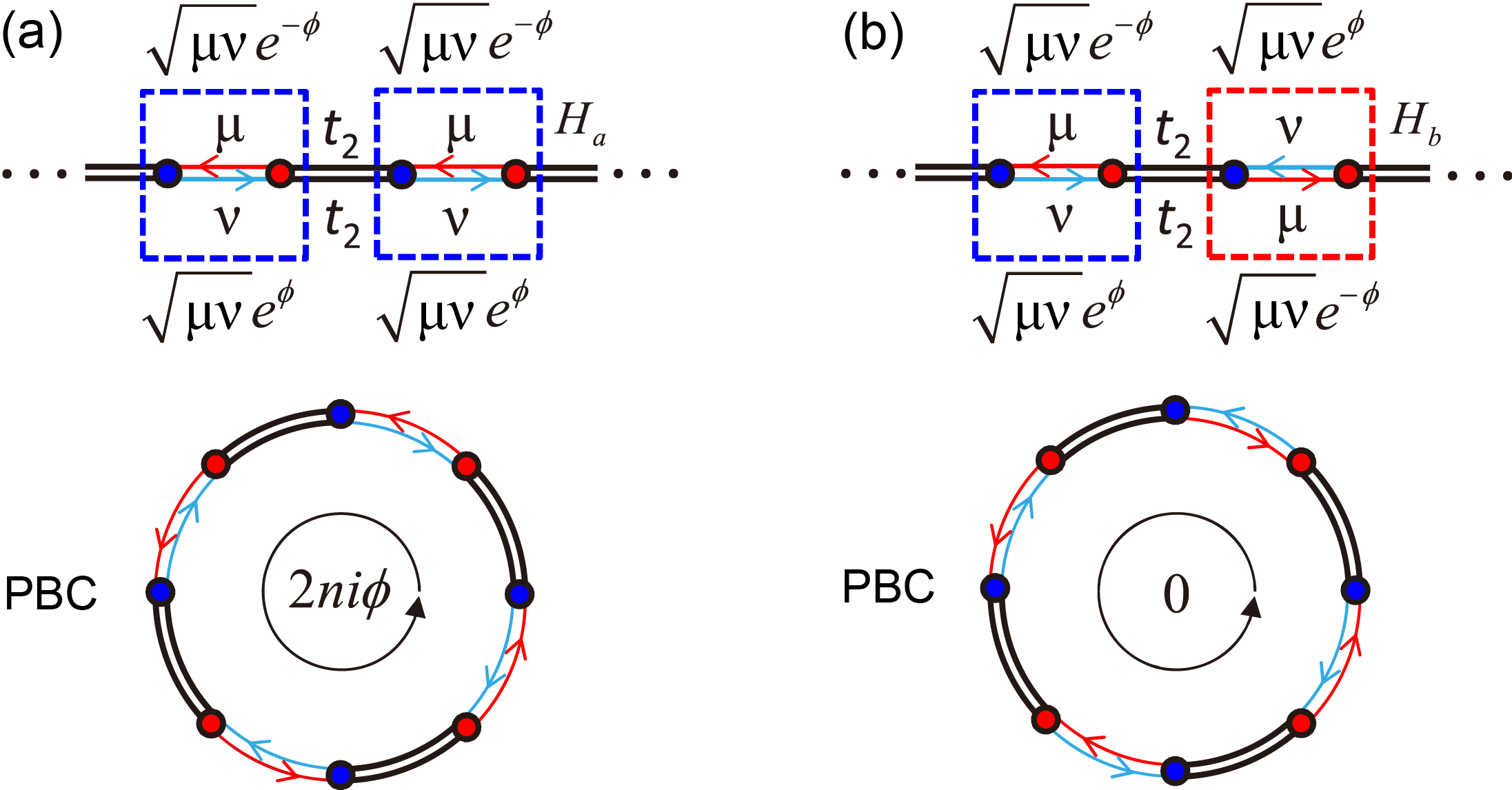}
\caption{Schematic of the imaginary gauge field and the magnetic flux.
In the present of imaginary gauge field, particles or photons circling one round accumulate an amplification and attenuation factor
$e^{\pm 2n\phi}$ ($0$) in opposite directions in system $a$ ($b$) under PBC; the enclosed imaginary magnetic flux is $2n i \phi$ ($0$). Under OBC, imaginary magnetic flux vanishes. The system size is $N=4n$.}
\label{figSI}
\end{figure}

In Fig. \ref{figSI}, we show the systems with asymmetric
coupling amplitudes are equivalent to systems with symmetric coupling
amplitudes that associate with amplification and attenuation factors for
tunneling in opposite directions. The amplification and attenuation factors $%
e^{\pm \phi }=e^{\pm i\Phi }$ with $\Phi =-i\phi $ indicates an imaginary
gauge field $\phi $. The imaginary gauge field enclosed in an area induces a
nonzero imaginary magnetic flux, which is a non-Hermitian Aharonov-Bohm (AB)
effect. For system $a$ under periodical boundary condition (PBC), $e^{\pm
\phi }$ accumulates along the translationally invariant direction; the
accumulated factor in one circle is $e^{\pm 2n\phi }$ for system size $N=4n$%
. The enclosed imaginary magnetic flux is $2n\left( i\phi \right) $
[Fig. \ref{figSI}(a)]. In contrast, the accumulated
factor $e^{\pm \phi }$ and $e^{\mp \phi }$ in the blue cell and the red cell
cancel each other in the compound four-site unit cell due to the
chiral-inversion symmetry; this indicates that the imaginary magnetic flux
does not exist in the chiral-inversion symmetric system $b$ under PBC
[Fig. \ref{figSI}(b)].

For $\mu \nu >0$, the Bloch Hamiltonian of system $a$ is%
\begin{equation}
H_{a}\left( k\right) =\left(
\begin{array}{cc}
0 & \mu +t_{2}e^{-ik} \\
\nu e^{\phi }+t_{2}e^{ik} & 0%
\end{array}%
\right) =\left(
\begin{array}{cc}
0 & \sqrt{\mu \nu }e^{-\phi }+t_{2}e^{-ik} \\
\sqrt{\mu \nu }e^{\phi }+t_{2}e^{ik} & 0%
\end{array}%
\right) ,
\end{equation}%
where $e^{-\phi }=\sqrt{\mu /\nu }$ and $\phi $ is an effective imaginary
gauge field. Under the basis $\{e^{-\phi }a_{k}^{\dagger },b_{k}^{\dagger
}\} $, we have $H_{a}\left( k\right) $ rewritten in the form of
\begin{equation}
H_{a}\left( k\right) =\left(
\begin{array}{cc}
0 & \sqrt{\mu \nu }+t_{2}e^{-i\left( k+i\phi \right) } \\
\sqrt{\mu \nu }+t_{2}e^{i\left( k+i\phi \right) } & 0%
\end{array}%
\right) .
\end{equation}%
The changing of basis indicates the implementation of a similar
transformation, which does not vary the energy spectrum. Then,
\begin{equation}
H_{a}\left( k\right) =[\sqrt{\mu \nu }+t_{2}\cos \left( k+i\phi \right)
]\sigma _{x}+t_{2}\sin \left( k+i\phi \right) \sigma _{y}.
\label{mumuLTzero0}
\end{equation}%
The energy bands are $E_{a,\pm }=\pm \sqrt{t_{2}^{2}+\mu \nu +2t_{2}\sqrt{%
\mu \nu }\cos \left( k+i\phi \right) }$.

For $\mu \nu <0$, the Bloch Hamiltonian of system $a$ can be similarly
rewritten in the form of symmetric coupling associated with an effective
imaginary gauge field that induces amplification/attenuation factor $e^{\pm
\phi }$. In the case of $\mu <0$ and $\nu >0$, we have $\mu =(i\sqrt{%
\left\vert \mu \nu \right\vert })(i\sqrt{\left\vert \mu /\nu \right\vert })$%
, $\nu =(i\sqrt{\left\vert \mu \nu \right\vert })(-i\sqrt{\left\vert \nu
/\mu \right\vert })$, thus we set $e^{-\phi }=i\sqrt{\left\vert \mu /\nu
\right\vert }$. In the case of $\mu >0$ and $\nu <0$, we have $\mu =(i\sqrt{%
\left\vert \mu \nu \right\vert })(-i\sqrt{\left\vert \mu /\nu \right\vert })$%
, $\nu =(i\sqrt{\left\vert \mu \nu \right\vert })(i\sqrt{\left\vert \nu /\mu
\right\vert })$, thus we set $e^{-\phi }=-i\sqrt{\left\vert \mu /\nu
\right\vert }$. In both cases, we obtain
\begin{equation}
H_{a}\left( k\right) =\left(
\begin{array}{cc}
0 & i\sqrt{\left\vert \mu \nu \right\vert }+t_{2}e^{-i\left( k+i\phi \right)
} \\
i\sqrt{\left\vert \mu \nu \right\vert }+t_{2}e^{i\left( k+i\phi \right) } & 0%
\end{array}%
\right) ,  \label{mumuLTzero}
\end{equation}%
under the basis $\{e^{-\phi }a_{k}^{\dagger },b_{k}^{\dagger }\}$.

Notably, under OBC, the imaginary gauge field is not enclosed; therefore,
the imaginary magnetic flux vanishes and the amplification/attenuation
factor $e^{\pm \phi }$ does not affect the system spectrum. Consequently,
systems $a$ and $b$ under OBC have identical spectra because the symmetric
coupling amplitude $\sqrt{\mu \nu }$ ($\mu \nu >0$) or $i\sqrt{\left\vert
\mu \nu \right\vert }$ ($\mu \nu <0$) without amplification/attenuation $%
e^{\pm \phi }$ in system $a$ is identical with that of system $b$. This
alternatively proves the identical spectra of systems $a$ and $b$ under OBC.
For instance, the Hamiltonians of a four-site systems $a$ and $b$ under OBC,
and the four-site Hamiltonian with symmetric coupling $\sqrt{\mu \nu }$ are
\begin{equation}
H_{a,4}=\left(
\begin{array}{cccc}
0 & \mu & 0 & 0 \\
\nu & 0 & t_{2} & 0 \\
0 & t_{2} & 0 & \mu \\
0 & 0 & \nu & 0%
\end{array}%
\right) ,H_{b,4}=\left(
\begin{array}{cccc}
0 & \mu & 0 & 0 \\
\nu & 0 & t_{2} & 0 \\
0 & t_{2} & 0 & \nu \\
0 & 0 & \mu & 0%
\end{array}%
\right) ,H_{S,4}=\left(
\begin{array}{cccc}
0 & \sqrt{\mu \nu } & 0 & 0 \\
\sqrt{\mu \nu } & 0 & t_{2} & 0 \\
0 & t_{2} & 0 & \sqrt{\mu \nu } \\
0 & 0 & \sqrt{\mu \nu } & 0%
\end{array}%
\right) .
\end{equation}%
We can easily check that they have identical eigenvalues of $E_{\pm ,\pm
}=\pm \left( t_{2}/2\right) \pm \sqrt{\left( t_{2}/2\right) ^{2}+\mu \nu }$.

The right eigenstate of $H_{a,4}$ ($H_{b,4}$) $\left\vert \psi
_{a,4}\right\rangle _{R}=[e^{-\phi }f_{1},f_{2},f_{3},e^{\phi }f_{4}]^{T}$ ($%
\left\vert \psi _{b,4}\right\rangle _{R}=[f_{1},e^{\phi }f_{2},e^{\phi
}f_{3},f_{4}]^{T}$) relates to the right eigenstate of $H_{S,4}$ $\left\vert
\psi _{S,4}\right\rangle _{R}=[f_{1},f_{2},f_{3},f_{4}]^{T}$ with the same
eigenvalue under a gauge transformation. The left eigenstate \cite{LFu} of $%
H_{a,4}$ ($H_{b,4}$) $_{L}\left\langle \psi _{a,4}\right\vert =[e^{\phi
}f_{1}^{\ast },f_{2}^{\ast },f_{3}^{\ast },e^{-\phi }f_{4}^{\ast }]$ ($%
_{L}\left\langle \psi _{b,4}\right\vert =[f_{1}^{\ast },e^{-\phi
}f_{2}^{\ast },e^{-\phi }f_{3}^{\ast },f_{4}^{\ast }]$) relates to the left
eigenstate of $H_{S,4}$ $_{L}\left\langle \psi _{S,4}\right\vert
=[f_{1}^{\ast },f_{2}^{\ast },f_{3}^{\ast },f_{4}^{\ast }]$ with the same
eigenvalue [$_{L}\left\langle \psi _{S,4}\right\vert =(\left\vert \psi
_{S,4}\right\rangle _{R})^{\dagger }$ in Hermitian systems]. The
biorthogonal norms of the eigenstates ($_{L}\langle \psi _{a,4}\left\vert
\psi _{a,4}\right\rangle _{R}=_{L}\langle \psi _{b,4}\left\vert \psi
_{b,4}\right\rangle _{R}=_{L}\langle \psi _{S,4}\left\vert \psi
_{S,4}\right\rangle _{R}$) are identical, but the Dirac norms ($_{R}\langle
\psi _{a,4}\left\vert \psi _{a,4}\right\rangle _{R}\neq _{R}\langle \psi
_{b,4}\left\vert \psi _{b,4}\right\rangle _{R}\neq _{R}\langle \psi
_{S,4}\left\vert \psi _{S,4}\right\rangle _{R}$) are different.

\subsection*{B. The inverse participation ratio of the bulk states}

\begin{figure}[thb]
\includegraphics[ bb=0 0 545 125, width=1.0\textwidth, clip]{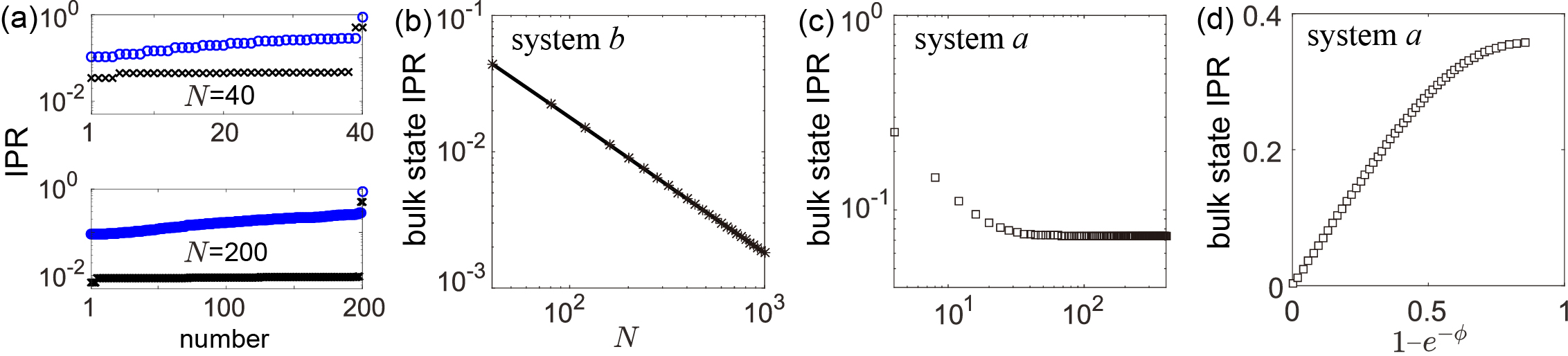}
\caption{(a) IPR of all the eigenstates, the blue circles (black
crosses) are for system $a$ ($b$). (b) Averaged IPR for all the bulk
states of system $b$, which scales as $1/N$. The system parameters are $t_1=1/4$,
$\gamma=1/2$, and $t_2=1$ in (a, b). (c) Averaged IPR for all the bulk
states of system $a$, $t_1=1$,
$\gamma=1/8$, and $t_2=1$. (d) Averaged IPR for all the bulk
states of system $a$ at large system size $N=400$ as a function of imaginary gauge field at $t_1=t_2=1$, $e^{-\phi }=\sqrt{\mu /\nu }$, $\mu=t_1-\gamma$, and $\nu=t_1+\gamma$.}
\label{figSScaling}
\end{figure}

The inverse participation ratio (IPR) $\sum_{j}|\psi _{j}|^{4}$ (with
normalization $\sum_{j}|\psi _{j}|^{2}=1$) for the eigenstates of system $a$
(blue circles) is system size insensitive compared with IPR of system $b$
(black crosses) in Fig.~\ref{figSScaling}(a). Fig.~\ref{figSScaling}(b) depicts the averaged IPR for the bulk
eigenstates of system $b$, which is inversely proportional to the system
size $\mathrm{IPR}\propto N^{-1}$; this reflects that all the bulk states of
system $b$ is extended state.

In Fig.~\ref{figSScaling}(c), the averaged IPR of system $a$
is depicted for weak non-Hermiticity. Correspondingly, the imaginary gauge
field strength is weak. It is noticed that at small system size, the bulk
states still exhibits the property of the extended states ($\mathrm{IPR}%
\propto N^{-1}$); in contrast, at large system size, the boundary
localization effect of bulk states appears, the averaged IPR is insensitive
to system size. The averaged IPR at large system size is a function of the
attenuation factor $e^{-\phi }$ [Fig.~\ref{figSScaling}(d)],
linearly increases as $\left( 1-e^{-\phi }\right) $ at weak non-Hermiticity
and is bounded at strong non-Hermiticity that leads to strong localization
of the bulk states at the system boundary.

\subsection*{C: Identical spectra for systems under open boundary condition}

The Hamiltonian of system $\rho $ ($\rho =a,b$) under open boundary
condition (OBC) is denoted as $H_{\rho ,N}$, where $N$ is the system size.
We have the matrices%
\begin{eqnarray}
H_{a,1}-EI_{1} &=&H_{b,1}-EI_{1}=\left( -E\right) ,  \label{S1} \\
H_{a,2}-EI_{2} &=&H_{b,2}-EI_{2}=\left(
\begin{array}{cc}
-E & \mu \\
\nu & -E%
\end{array}%
\right) ,  \label{S2}
\end{eqnarray}%
and%
\begin{equation}
H_{a,N}-EI_{N}=\left(
\begin{array}{cccccc}
-E & \mu &  &  &  &  \\
\nu & -E & t_{2} &  &  &  \\
& t_{2} & -E & \mu &  &  \\
&  & \nu & -E & t_{2} &  \\
&  &  & t_{2} & \ddots & \ddots \\
&  &  &  & \ddots & \ddots%
\end{array}%
\right) ,H_{b,N}-EI_{N}=\left(
\begin{array}{cccccc}
-E & \mu &  &  &  &  \\
\nu & -E & t_{2} &  &  &  \\
& t_{2} & -E & \nu &  &  \\
&  & \mu & -E & t_{2} &  \\
&  &  & t_{2} & \ddots & \ddots \\
&  &  &  & \ddots & \ddots%
\end{array}%
\right) ,
\end{equation}%
where $I_{N}$ is the $N\times N$ identical matrix. The determinant $%
D_{N}\left( \rho \right) $ for system $\rho $ is
\begin{equation}
D_{N}\left( \rho \right) =\mathrm{Det}\left( H_{\rho ,N}-EI_{N}\right) .
\end{equation}%
They satisfy a recursion relationship%
\begin{eqnarray}
D_{2m-1}\left( \rho \right) &=&\left( -E\right) D_{2m-2}\left( \rho \right)
-t_{2}^{2}D_{2m-3}\left( \rho \right) , \\
D_{2m}\left( \rho \right) &=&\left( -E\right) D_{2m-1}\left( \rho \right)
-\mu \nu D_{2m-2}\left( \rho \right) ,
\end{eqnarray}%
for integer $m$ from $2$ to $2n$. Equations (\ref{S1})-(\ref{S2}) yield $%
D_{1}\left( a\right) =D_{1}\left( b\right) $ and $D_{2}\left( a\right)
=D_{2}\left( b\right) $, we acquire $D_{3}\left( a\right) =D_{3}\left(
b\right) $ and consequently $D_{N}\left( a\right) =D_{N}\left( b\right) $.
The eigenvalues $E$ of $H_{\rho ,N}$ for system $\rho $ is obtained from $%
D_{N}\left( \rho \right) =0$; therefore, two Hamiltonians $H_{a}$ and $H_{b}$
under OBC possess identical spectra.

\subsection*{D. Chiral-inversion symmetric systems}

\begin{figure}[thb]
\includegraphics[ bb=0 0 280 280, width=0.5\textwidth, clip]{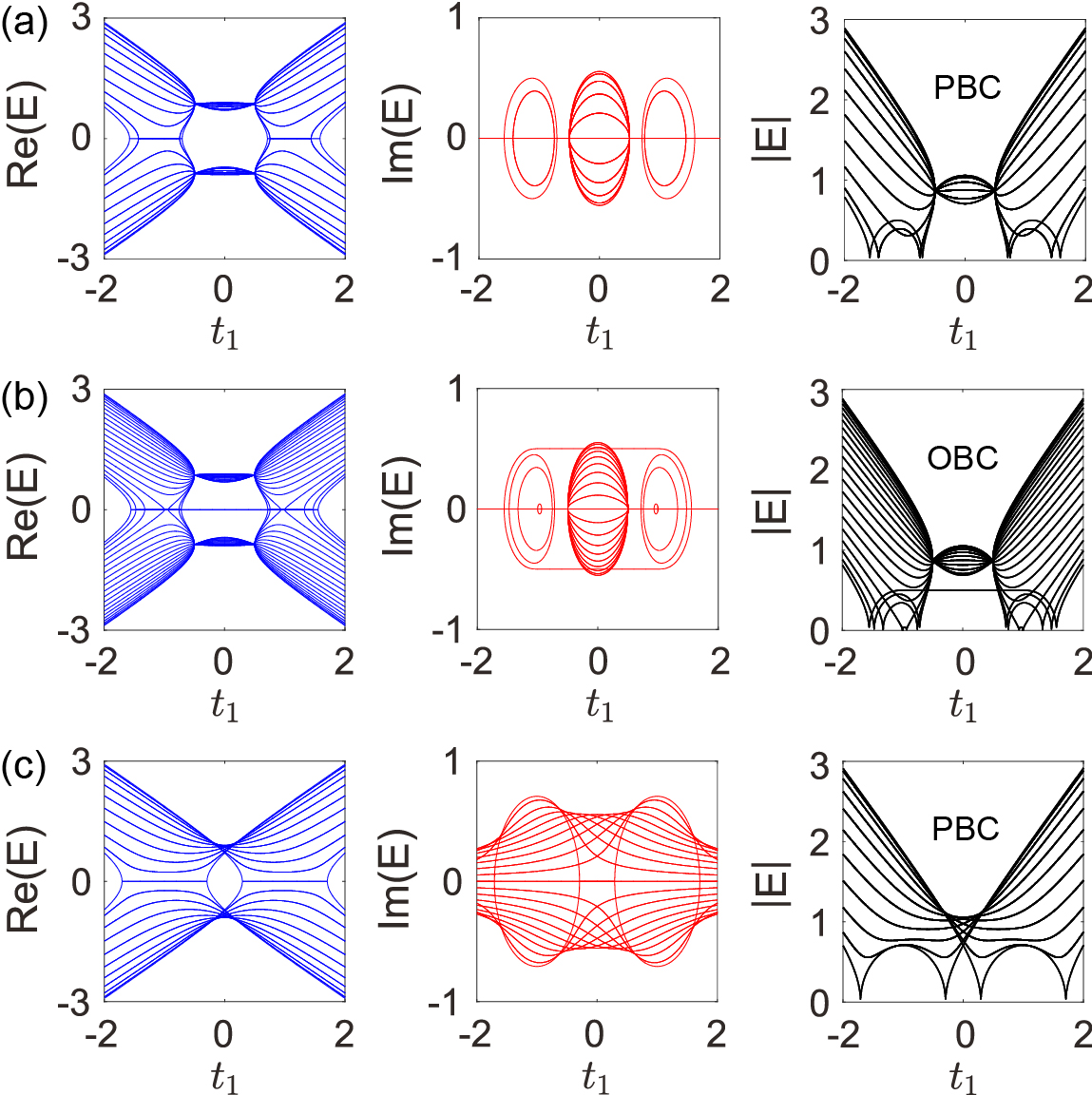}
\caption{Energy spectrum for system $b$ with staggered gain and loss $\left\{ i\Gamma ,-i\Gamma ,i\Gamma ,-i\Gamma \right\} $ under (a)
PBC and (b) OBC. (c) Energy spectrum for system $a$ with staggered gain and loss $\left\{ i\Gamma ,-i\Gamma \right\} $ under
PBC. The corresponding OBC spectrum is in (b). The parameters are $N=40$,
$\gamma=1/2$, $\Gamma=1/2$, and $t_2=1$.} \label{figS3}
\end{figure}

System $b^{\prime }$ is system $b$ with the asymmetric couplings $\left( \nu
,\mu \right) $ in the red unit cell (chiral-inversion symmetric system $%
H_{b} $ in Fig. 1 of the main paper) with an additional minus sign in front [%
$\left( \nu ,\mu \right) $ substituted by $\left( -\nu ,-\mu \right) $]; the
Bloch Hamiltonian of system $b^{\prime }$ reads%
\begin{equation}
H_{b^{\prime }}\left( k\right) =\left(
\begin{array}{cccc}
0 & \mu & 0 & t_{2}e^{-ik} \\
\nu & 0 & t_{2} & 0 \\
0 & t_{2} & 0 & -\nu \\
t_{2}e^{ik} & 0 & -\mu & 0%
\end{array}%
\right) .
\end{equation}%
For even $n$, systems $b$ and $b^{\prime }$ in the real-space are connected
through a similar transformation
\begin{equation}
U_{b^{\prime }b}=P_{n/2}\otimes \sigma _{y}\otimes \left[ \left( \sigma
_{x}+i\sigma _{y}\right) \otimes \left( i\sigma _{y}\right) +\left( i\sigma
_{y}-\sigma _{x}\right) \otimes \sigma _{x}\right] /2,
\end{equation}%
that $U_{b^{\prime }b}H_{b^{\prime }}U_{b^{\prime }b}^{\dagger }=H_{b}$.
Therefore, systems $b$ and $b^{\prime }$ have identical band structures and
topological properties. System $b^{\prime }$ has the combined
chiral-inversion symmetry, $\left( \mathcal{SP}\right) H_{b^{\prime }}\left(
k\right) \left( \mathcal{SP}\right) ^{-1}=-H_{b^{\prime }}\left( -k\right) $%
, where $\mathcal{SP}=\sigma _{y}\otimes \sigma _{x}$; and $U_{\mathcal{SP}%
}H_{b^{\prime }}U_{\mathcal{SP}}^{\dagger }=-H_{b^{\prime }}$, where $U_{%
\mathcal{SP}}=P_{n}\otimes \left( \sigma _{y}\otimes \sigma _{x}\right) $.
The chiral symmetry is satisfied, $\mathcal{S}H_{b^{\prime }}\mathcal{S}%
^{-1}=-H_{b^{\prime }}$, where $\mathcal{S=}I_{2n}\otimes \sigma _{z}$; but
the inversion symmetry is violated.

System $b$ with alternative on-site gain and loss $\left\{ i\Gamma ,-i\Gamma
,i\Gamma ,-i\Gamma \right\} $ introduced in the four-site unit cell
possesses the chiral-inversion symmetry that $\left( \mathcal{SP}\right)
H_{b}\left( k\right) \left( \mathcal{SP}\right) ^{-1}=-H_{b}\left( -k\right)
$, where $\mathcal{SP}=\sigma _{x}\otimes \sigma _{y}$; and $U_{\mathcal{SP}%
}H_{b}U_{\mathcal{SP}}^{\dagger }=-H_{b}$, where $U_{\mathcal{SP}%
}=P_{2n}\otimes \sigma _{y}$. Notably, both the chiral symmetry and the
inversion symmetry are violated. The energy spectra under PBC and OBC are
depicted in Figs. \ref{figS3}(a) and \ref{figS3}(b),
respectively.

Their energy spectra under OBC and PBC are in Figs. \ref%
{figS3}(b) and \ref{figS3}(c) for system $a$ with $\left\{ i\Gamma ,-i\Gamma
\right\} $ introduced in the unit cell, which has identical (different)
spectrum under OBC (PBC) with that of system $b$ composed by introducing $%
\left\{ i\Gamma ,-i\Gamma ,i\Gamma ,-i\Gamma \right\} $ aiming to fix the
chiral-inversion symmetry in system $a$. Notably, the band touching
degeneracy (exceptional) points in the energy spectra of the systems under
PBC and OBC are in accord with each other; the system boundary does not
alter most of the bulk states; the bulk states are all extended states; and
the non-Hermitian skin effect is absent.

\subsection*{E. Topological characterization}

The bulk-boundary correspondence is valid for the chiral-inversion symmetric
system $b$ due to the absence of nonzero imaginary magnetic flux. The bulk
Bloch Hamiltonian of system $b$ characterizes the topological properties of
system $b$ and the (non)existence of edge states under OBC; the topological
invariant is a Bloch topological invariant for system $b$. The bulk Bloch
Hamiltonian of system $b$ also characterizes the topological properties of
system $a$ under OBC because the identical spectra of systems $b$ and $a$
under OBC. However, the bulk-boundary correspondence is invalid due to the
lack of chiral-inversion symmetry in system $a$; the bulk Bloch Hamiltonian
of system $a$ is not able to characterize the (non)existence of edge states
in system $a$ under OBC. Thus, the topological invariant is a non-Bloch
topological invariant for system $a$.

In system $b$, the bulk topological properties relate to the (non)existence
of edge states. Here we calculate the bulk topological invariant of system $%
b $, which is capable of characterizing the topologies of both systems $a$
and $b$ under OBC. The Bloch Hamiltonian of system $b$ is a $4\times 4$
matrix; after a similar transformation, the Bloch Hamiltonian can be
expressed in the form of $\vec{B}\cdot \vec{\sigma}$ with a two-site unit
cell, then we define a vector field $\mathbf{F}\left( \mathbf{k}\right) $
that is associated with the Bloch Hamiltonian. The topological defects with
vortices or antivortices in the vector field indicate the phase transition
points. The vorticity of the topological defects is a topological invariant.
We consider that $t_{1}$ and $\gamma $ are real numbers, $\mu =t_{1}-\gamma $
and $\nu =t_{1}+\gamma $, discussions on other cases are similarly following
the same procedure below.

For $\mu \nu >0$, $\mu $, $\nu $, and $\sqrt{\mu \nu }$ are positive real
numbers. The Bloch Hamiltonian $H_{b}\left( k\right) $ under a similar
transformation $U_{\mu \nu >0}=\mathrm{diag}\left( \sqrt{\nu },\sqrt{\mu },%
\sqrt{\mu },\sqrt{\nu }\right) $ that only consists of diagonal elements,
yields%
\begin{equation}
U_{\mu \nu >0}H_{b}\left( k\right) U_{\mu \nu >0}^{-1}=\left(
\begin{array}{cccc}
0 & \sqrt{\mu \nu } & 0 & t_{2}e^{-ik} \\
\sqrt{\mu \nu } & 0 & t_{2} & 0 \\
0 & t_{2} & 0 & \sqrt{\mu \nu } \\
t_{2}e^{ik} & 0 & \sqrt{\mu \nu } & 0%
\end{array}%
\right) ,  \label{Hermitian}
\end{equation}%
which equals to the Bloch Hamiltonian of chiral-inversion symmetric system $%
b $ with the imaginary gauge field $\pm i\phi $ removed. The eigenvalues are
symmetric $E_{b,\pm ,\pm }=\pm \sqrt{t_{2}^{2}+\mu \nu \pm 2t_{2}\sqrt{\mu
\nu }\cos (k/2)}$. In the discrete system with lattice size $N=4n$, the wave
vector $k$ is $k=2\pi m/n$, $m\in \lbrack 1,n]$ ($m,n$ are positive
integers) for the Bloch Hamiltonian with a four-site unit cell. The Bloch
Hamiltonian of equation (\ref{Hermitian})\ also equals to system $a$ with
all the asymmetric couplings $\mu $ and $\nu $ replaced by the symmetric
coupling $\sqrt{\mu \nu }$ and taken two unit cells as a compound unit cell.
The Bloch Hamiltonian of equation (\ref{Hermitian}) is rewritten in the form
of%
\begin{equation}
h_{b}\left( k\right) =\left(
\begin{array}{cc}
0 & \sqrt{\mu \nu }+t_{2}e^{-ik} \\
\sqrt{\mu \nu }+t_{2}e^{ik} & 0%
\end{array}%
\right) .  \label{h_Hermitian}
\end{equation}%
$h_{b}\left( k\right) =\vec{B}\cdot \vec{\sigma}$, where the effective
magnetic field is
\begin{equation}
\vec{B}=\left( \sqrt{\mu \nu }+t_{2}\cos k,t_{2}\sin k,0\right) .
\end{equation}%
Notably, the wave vector $k$ is $k=\pi m/n$, $m\in \lbrack 1,2n]$ for the
Bloch Hamiltonian with a two-site unit cell.

We define a vector field $\mathbf{F}\left( \mathbf{k}\right) =\left(
\left\langle \sigma _{x}\right\rangle ,\left\langle \sigma _{y}\right\rangle
\right) $ to characterize the topology of $h_{b}\left( k\right) $. The
eigenstates associated with $E_{\pm }\left( k\right) =\pm \sqrt{\left( \sqrt{%
\mu \nu }+t_{2}e^{-ik}\right) \left( \sqrt{\mu \nu }+t_{2}e^{ik}\right) }$
are
\begin{equation}
\psi _{\pm }\left( k\right) =\frac{1}{\sqrt{2\sqrt{\mu \nu +2\sqrt{\mu \nu }%
t_{2}\cos k+t_{2}^{2}}}}\left(
\begin{array}{c}
\pm \sqrt{\sqrt{\mu \nu }+t_{2}e^{-ik}} \\
\sqrt{\sqrt{\mu \nu }+t_{2}e^{ik}}%
\end{array}%
\right) .
\end{equation}%
The average values of the Pauli matrices associated with the two-component
effective magnetic field $\left\langle \sigma _{x,y}\right\rangle _{\pm
}=\left\langle \psi _{\pm }\left( k\right) \right\vert \sigma
_{x,y}\left\vert \psi _{\pm }\left( k\right) \right\rangle $ are%
\begin{equation}
\left\langle \sigma _{x}\right\rangle _{\pm }=\frac{\pm \left( \sqrt{\mu \nu
}+t_{2}\cos k\right) }{\sqrt{\mu \nu +2\sqrt{\mu \nu }t_{2}\cos k+t_{2}^{2}}}%
,\left\langle \sigma _{y}\right\rangle _{\pm }=\frac{\pm t_{2}\sin k}{\sqrt{%
\mu \nu +2\sqrt{\mu \nu }t_{2}\cos k+t_{2}^{2}}},
\end{equation}%
i.e., $(\left\langle \sigma _{x}\right\rangle _{\pm },\left\langle \sigma
_{y}\right\rangle _{\pm })=\left( B_{x},B_{y}\right) /E_{\pm }$; thus, $%
(\left\langle \sigma _{x}\right\rangle _{\pm },\left\langle \sigma
_{y}\right\rangle _{\pm })$ reflects the topological properties of the Bloch
bands and the system. The vector field $\mathbf{F}\left( \mathbf{k}\right) $
under either eigenstate yields the same winding number $w=\oint_{L}\left(
2\pi \right) ^{-1}(\hat{F}_{x}\nabla \hat{F}_{y}-\hat{F}_{y}\nabla \hat{F}%
_{x})\mathrm{d}\mathbf{k}$ in the parameter plane $\mathbf{k=}\left(
k,t_{2}\right) $, where $\hat{F}_{x(y)}=F_{x(y)}/\sqrt{F_{x}^{2}+F_{y}^{2}}$
and $\nabla =\partial /\partial \mathbf{k}$.$\ $The phase transition occurs
at $\left( k,t_{2}\right) =\left( 0,-\sqrt{\mu \nu }\right) $ or $\left( \pm
\pi ,\sqrt{\mu \nu }\right) $, which are the band touching degeneracy
points. They are topological defects in the vector field possessing integer
topological charges (vortices and antivortices) as depicted in Fig. 3(a) in
the main paper. The winding number $w$ characterizing the vorticity of the
topological defects, is a topological invariant.

For $\mu \nu <0$, $-\mu $, $\nu $, and $\sqrt{-\mu \nu }$ are positive real
numbers. The Bloch Hamiltonian $H_{b}\left( k\right) $ under a similar
transformation $U_{\mu \nu <0}=\mathrm{diag}\left( \sqrt{\nu },i\sqrt{-\mu }%
,i\sqrt{-\mu },\sqrt{\nu }\right) $, yields%
\begin{equation}
U_{\mu \nu <0}H_{b}\left( k\right) U_{\mu \nu <0}^{-1}=\left(
\begin{array}{cccc}
0 & i\sqrt{-\mu \nu } & 0 & t_{2}e^{-ik} \\
i\sqrt{-\mu \nu } & 0 & t_{2} & 0 \\
0 & t_{2} & 0 & i\sqrt{-\mu \nu } \\
t_{2}e^{ik} & 0 & i\sqrt{-\mu \nu } & 0%
\end{array}%
\right) ,  \label{nonHermitian}
\end{equation}%
which is the Bloch Hamiltonian of system $a$ with all the asymmetric
couplings $\mu $ and $\nu $ replaced by the symmetric coupling $i\sqrt{-\mu
\nu }$ and taken two unit cells as a compound unit cell. The Bloch
Hamiltonian can be rewritten as%
\begin{equation}
h_{b}\left( k\right) =\left(
\begin{array}{cc}
0 & i\sqrt{-\mu \nu }+t_{2}e^{-ik} \\
i\sqrt{-\mu \nu }+t_{2}e^{ik} & 0%
\end{array}%
\right) ,
\end{equation}%
where the effective magnetic field is
\begin{equation}
\vec{B}=\left( i\sqrt{-\mu \nu }+t_{2}\cos k,t_{2}\sin k,0\right) .
\end{equation}%
the eigenvalues are $E_{\pm }\left( k\right) =\pm \sqrt{\left( i\sqrt{-\mu
\nu }+t_{2}e^{-ik}\right) \left( i\sqrt{-\mu \nu }+t_{2}e^{ik}\right) }$;
correspondingly, the eigenstates are
\begin{equation}
\psi _{\pm }\left( k\right) =\frac{1}{\sqrt{\Delta }}\left(
\begin{array}{c}
\pm \sqrt{i\sqrt{-\mu \nu }+t_{2}e^{-ik}} \\
\sqrt{i\sqrt{-\mu \nu }+t_{2}e^{ik}}%
\end{array}%
\right) ,
\end{equation}%
where $\Delta =\sqrt{t_{2}^{2}-2\sqrt{-\mu \nu }t_{2}\sin k-\mu \nu }+\sqrt{%
t_{2}^{2}+2\sqrt{-\mu \nu }t_{2}\sin k-\mu \nu }$.

The average values of $\left\langle \sigma _{x,y}\right\rangle _{\pm
}=\left\langle \psi _{\pm }\left( k\right) \right\vert \sigma
_{x,y}\left\vert \psi _{\pm }\left( k\right) \right\rangle $ are%
\begin{equation}
\left\langle \sigma _{x}\right\rangle _{\pm }=\frac{\pm \left( \sqrt{%
e^{-2ik}t_{2}^{2}-\mu \nu }+\sqrt{e^{2ik}t_{2}^{2}-\mu \nu }\right) }{\Delta
},\left\langle \sigma _{y}\right\rangle _{\pm }=\frac{\pm i\left( \sqrt{%
e^{-2ik}t_{2}^{2}-\mu \nu }-\sqrt{e^{2ik}t_{2}^{2}-\mu \nu }\right) }{\Delta
}.
\end{equation}%
The phase transition points are $\left( k,t_{2}\right) =\left( \pm \pi
/2,\pm \sqrt{-\mu \nu }\right) $, which are the band touching EPs. They are
topological defects in the vector field possessing half-integer topological
charges (vortices and antivortices) as depicted in Fig. 3(b) in the main
paper.

\begin{figure}[tbh]
\includegraphics[ bb=0 0 525 120, width=1.0\textwidth, clip]{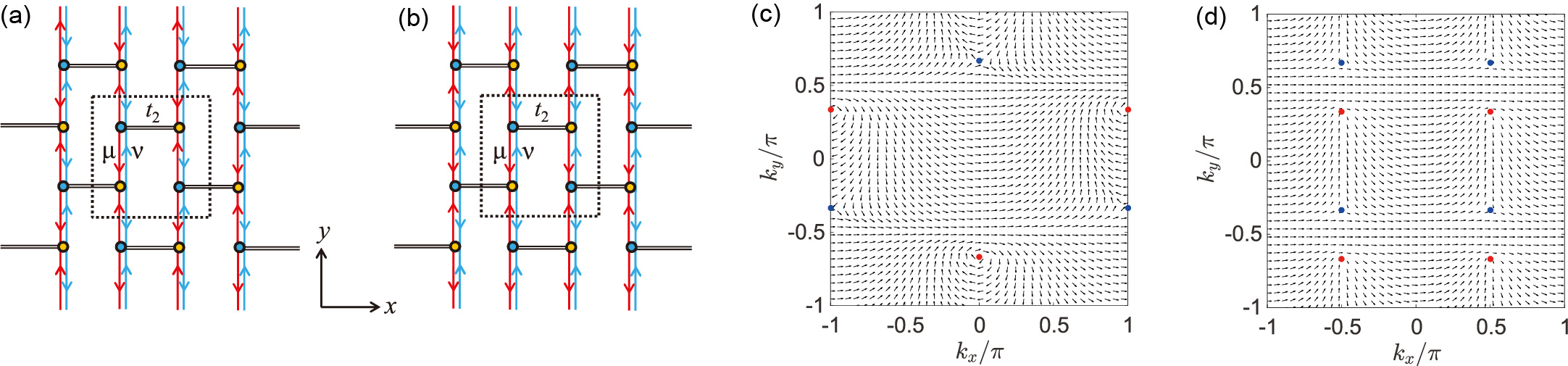}
\caption{(a, b) Schematic of the 2D brick wall lattice. The 2D lattices in the $x$-direction are the 1D non-Hermitian SSH lattices in Fig 1 of the main paper,
respectively. (c, d) The vector field $\mathbf{F}\left( k_{x},k_{y}\right) $ associated with the $E_{b,+}^{(2D)}$ state. The parameters are $t_2=1$ and (c) $\mu \nu=1$, (d) $\mu \nu=-1$.}
\label{fig2D}
\end{figure}

The two-dimensional (2D) brick wall lattices are schematically illustrated
in Figs. \ref{fig2D}(a) and \ref{fig2D}(b), along one
direction of the 2D lattices are systems $a$ and $b$, respectively. In the
momentum space, the Bloch Hamiltonians $H_{a}^{(2D)}\left( k\right) $ and $%
H_{b}^{(2D)}\left( k\right) $ are%
\begin{eqnarray}
H_{a}^{(2D)}\left( k\right) &=&\left(
\begin{array}{cccc}
0 & 2\mu \cos \left( k_{y}\right) & 0 & t_{2}e^{-ik_{x}} \\
2\nu \cos \left( k_{y}\right) & 0 & t_{2}e^{ik_{x}} & 0 \\
0 & t_{2}e^{-ik_{x}} & 0 & 2\mu \cos \left( k_{y}\right) \\
t_{2}e^{ik_{x}} & 0 & 2\nu \cos \left( k_{y}\right) & 0%
\end{array}%
\right) , \\
H_{b}^{(2D)}\left( k\right) &=&\left(
\begin{array}{cccc}
0 & 2\mu \cos \left( k_{y}\right) & 0 & t_{2}e^{-ik_{x}} \\
2\nu \cos \left( k_{y}\right) & 0 & t_{2}e^{ik_{x}} & 0 \\
0 & t_{2}e^{-ik_{x}} & 0 & 2\nu \cos \left( k_{y}\right) \\
t_{2}e^{ik_{x}} & 0 & 2\mu \cos \left( k_{y}\right) & 0%
\end{array}%
\right) .
\end{eqnarray}%
The vortices and antivortices associated with the phase transition points in
the Brillouin zone in the $k_{x}-k_{y}$ space are shown in Figs. \ref{fig2D}(c) and \ref{fig2D}(d), which are in accord with that
revealed in Figs. 3(a) and 3(b) in the main paper.

Under a similar transformation $U_{b}^{(2D)}=\mathrm{diag}\left( \sqrt{\nu },%
\sqrt{\mu },\sqrt{\mu },\sqrt{\nu }\right) $, the Bloch Hamiltonian $%
H_{b}^{(2D)}\left( k\right) $ changes into
\begin{equation}
U_{b}^{(2D)}H_{b}^{(2D)}\left( k\right) U_{b}^{(2D)-1}=\left(
\begin{array}{cccc}
0 & 2\sqrt{\mu \nu }\cos k_{y} & 0 & t_{2}e^{-ik_{x}} \\
2\sqrt{\mu \nu }\cos k_{y} & 0 & t_{2}e^{ik_{x}} & 0 \\
0 & t_{2}e^{-ik_{x}} & 0 & 2\sqrt{\mu \nu }\cos k_{y} \\
t_{2}e^{ik_{x}} & 0 & 2\sqrt{\mu \nu }\cos k_{y} & 0%
\end{array}%
\right) ,
\end{equation}%
and the corresponding two-site unit cell Bloch Hamiltonian is%
\begin{equation}
h_{b}^{(2D)}=\left(
\begin{array}{cc}
0 & 2\sqrt{\mu \nu }\cos k_{y}+t_{2}e^{-ik_{x}} \\
2\sqrt{\mu \nu }\cos k_{y}+t_{2}e^{ik_{x}} & 0%
\end{array}%
\right) ,
\end{equation}%
where the eigenvalues are $E_{b,\pm }^{(2D)}=\pm \sqrt{t_{2}^{2}+4\sqrt{\mu
\nu }\cos k_{x}\cos k_{y}+4\mu \nu \cos ^{2}k_{y}}$.

\subsection*{F. Energy spectra for complex asymmetric coupling}

\begin{figure}[th]
\includegraphics[ bb=0 0 580 90, width=1.0\textwidth, clip]{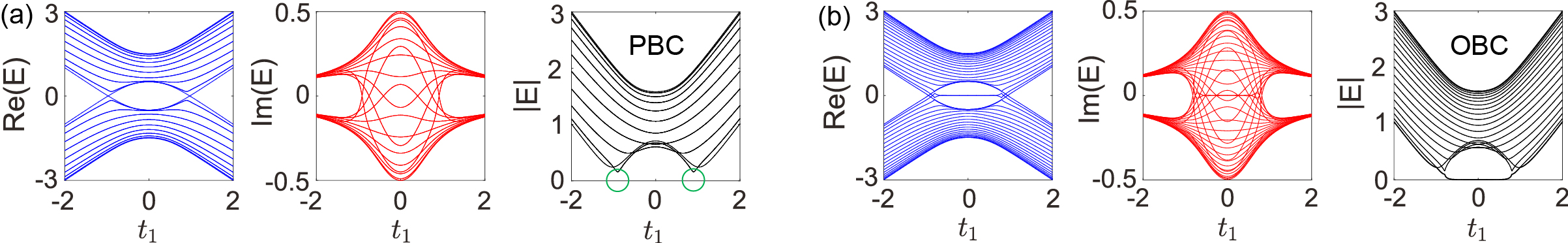}
\caption{Energy spectra of system $b$ under (a) PBC and (b) OBC.
The parameters are $N=40$, $\gamma=\sqrt{1/2}e^{i\pi/4}$, and $t_2=1$.} %
\label{figS2}
\end{figure}

For system $b$ at $\gamma =\sqrt{1/2}e^{i\pi /4}$, the phase transition
points are $t_{1}=\pm \sqrt[4]{3/4}\approx 0.93$ and $\left\vert \cos \left(
k\right) \right\vert =\sqrt{(2+\sqrt{3})}/2$. The band touching degeneracy
points may not be seen in the discrete system with small system size due to
the finite number of discrete $k$. The energy spectra for $\gamma =\sqrt{1/2}%
e^{i\pi /4}$ are depicted in Fig. \ref{figS2} for $N=40$.\

For system $b$ at complex $\gamma $, the momenta $k$ for band touching in
the energy spectra are no longer $0$, $\pm \pi /2$, or $\pm \pi $ and may
not be seen in the discrete system with small size. The nonvanishing gap in $%
|E|$ diminishes (vanishes) as system size increasing ($N\rightarrow \infty $%
). The nonvanishing gap in $|E|$ shown inside the green circles in
Fig. \ref{figS2}(a) is a finite size effect of the discrete
system; as $N$ increases, the gap vanishes and the band touching degeneracy
points reveal.

\subsection*{G. Edge states for systems with a defective unit cell at
boundary}

\begin{figure}[tbh]
\includegraphics[ bb=0 0 580 90, width=1.0\textwidth, clip]{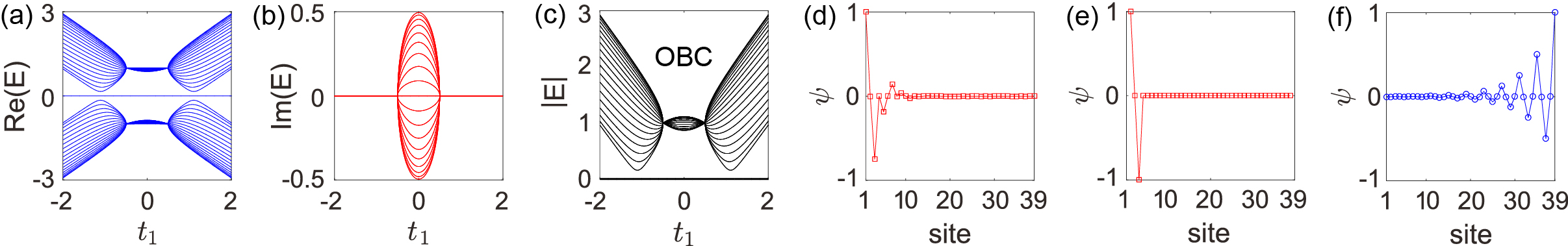}
\caption{(a-c) Energy spectra of system $b$ with $N=39$ at parameters
$\gamma=1/2$ and $t_2=1$. The zero edge state at (d) $t_1=1/4$, (e)
$t_1=1/2$, and (f) $t_1=3/2$.} \label{figS1}
\end{figure}

For system $b$ with an even site number (total site number $N=4n-2$), we
consider that $H_{b}$ under OBC with two sites (inside the red rectangle of
chiral-inversion symmetric system $H_{b}$ in Fig. 1 of the main paper) at
the right boundary are missing. Two zero edge states localize at the left
and the right boundaries, respectively. The left edge state is Eq. (5) in
the main paper. The right edge state localized at the right boundary is $%
\psi _{2j-1}=0$ and
\begin{equation}
\psi _{N-2j}=-[\left( \nu +\mu \right) +\left( -1\right) ^{j}\left( \nu -\mu
\right) ]/\left( 2t_{2}\right) \psi _{N+2-2j},
\end{equation}%
at large system size limit ($N\gg 1$). For the anomalous edge states at the
EPs ($t_{1}^{2}=\gamma ^{2}$), they are localized at a single unit cell at
system boundary. For $t_{1}=-\gamma $, the left (right) edge state is $\psi
_{1}=1$ ($\psi _{N}=1$); and for $t_{1}=\gamma $, the right edge state is $%
\psi _{N}=1$; the left edge state is $\psi _{1}=-\left( +\right) \psi _{3}=1$
for $t_{1}/t_{2}>0$ ($t_{1}/t_{2}<0$).

For system $b$ with an odd site number, the energy spectra are depicted in
Figs. \ref{figS1}(a)$-$\ref{figS1}(c), the edge state is
depicted in Figs. \ref{figS1}(d)$-$\ref{figS1}(f). Only one
zero edge state exists in this situation. If the unit cell at the right
boundary is defective, in the situation that $\left\vert \mu \nu \right\vert
<t_{2}^{2}$, the edge state localizes at the left boundary, the wave
function is Eq. (5) in the main paper; in the situation that $\left\vert \mu
\nu \right\vert >t_{2}^{2}$, for the system with site number $N=4n-1$, the
edge state localized at the right boundary is $\psi _{2j}=0$ and
\begin{equation}
\psi _{N-2j}=-\left( 2t_{2}\right) /[\left( \mu +\nu \right) +\left(
-1\right) ^{j}\left( \mu -\nu \right) ]\psi _{N+2-2j};
\end{equation}%
for the system with site number $N=4n-3$, the edge state localized at the
right boundary is $\psi _{2j}=0$ and%
\begin{equation}
\psi _{N-2j}=-\left( 2t_{2}\right) /[\left( \nu +\mu \right) +\left(
-1\right) ^{j}\left( \nu -\mu \right) ]\psi _{N+2-2j}.
\end{equation}%
For the anomalous edge states at the EPs ($t_{1}^{2}=\gamma ^{2}$) and the
systems with site numbers $N=4n-1$ and $4n-3$, the left edge state is $\psi
_{1}=-\left( +\right) \psi _{3}=1$ for $t_{1}/t_{2}>0$ ($t_{1}/t_{2}<0$) at $%
t_{1}=\gamma $ and $\psi _{1}=1$ at $t_{1}=-\gamma $.

In contrast, for system $a$ with an odd site number having a defective unit
cell at the right boundary, only one zero state exists. The right boundary
state is $\psi _{2j}=0$ and $\psi _{N-2j}=\left( -t_{2}/\nu \right) \psi
_{N+2-2j}$ when $\left\vert t_{2}\right\vert <\sqrt{\mu \nu }$. At the EPs,
the zero state localizes at one site on the left boundary $\psi _{1}=1$ for $%
t_{1}=-\gamma $; and the zero state is extended, being $\psi _{2j}=0$ and $%
\psi _{2j-1}=-\left( +\right) \psi _{2j+1}$ for $t_{1}=\gamma $ at $%
t_{1}/t_{2}>0$ ($t_{1}/t_{2}<0$).

\subsection*{H. Equivalence between systems and their connections}

\begin{figure}[tb]
\includegraphics[ bb= 0 0 530 60, width=0.98\textwidth, clip]{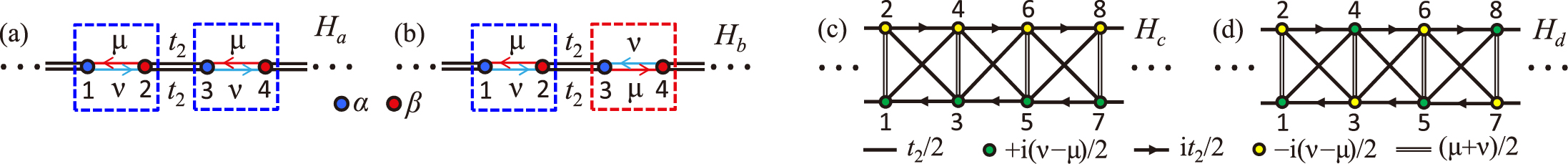}
\caption{(a, b) Schematic of the 1D SSH lattices, the unit cells are
indicated in the squares. (c) and (d) are the equivalent non-Hermitian
Creutz ladders of (a) and (b), respectively. $H_{\rho}(k)$ ($H_{\rho}$) is
the Hamiltonian $\rho$ in the $k$-space (real-space), where $\rho=a, b, c,
d$. The system sizes are all $N=4n$. This indicates that the imaginary gauge field relates to the real gauge field associated with balanced gain and loss.}
\label{figSillus}
\end{figure}

The systems discussed have the total lattice size $N=4n$, and the labels are
marked in Fig. \ref{figSillus}. In the real-space, through a
unitary transformation%
\begin{equation}
U=\frac{1}{\sqrt{2}}I_{2n}\otimes \left( i\sigma _{x}+I_{2}\right) ,
\label{UnitaryTransformation}
\end{equation}%
the systems $a$ and $b$ change into their corresponding quasi-1D Creutz
ladder systems $c$ and $d$, respectively; i.e., $H_{c(d)}=UH_{a(b)}U^{-1}$;
where $\otimes $ is the Kronecker product and $I_{2n}$ is the $2n\times 2n$
identical matrix. After the $U$ transformation, the SSH chain with
asymmetric coupling changes into the Creutz ladder in Fig. 4 of the main
paper.

In the $k$-space, the Bloch Hamiltonians are%
\begin{equation}
H_{a}\left( k\right) =\left(
\begin{array}{cc}
0 & \mu +t_{2}e^{-ik} \\
\nu +t_{2}e^{ik} & 0%
\end{array}%
\right) ,H_{c}\left( k\right) =\frac{1}{2}\left(
\begin{array}{cc}
i\nu -i\mu -2t_{2}\sin k & \mu +\nu +2t_{2}\cos k \\
\mu +\nu +2t_{2}\cos k & i\mu -i\nu +2t_{2}\sin k%
\end{array}%
\right) ,
\end{equation}%
\begin{equation}
H_{b}\left( k\right) =\left(
\begin{array}{cccc}
0 & \mu & 0 & t_{2}e^{-ik} \\
\nu & 0 & t_{2} & 0 \\
0 & t_{2} & 0 & \nu \\
t_{2}e^{ik} & 0 & \mu & 0%
\end{array}%
\right) ,H_{d}\left( k\right) =\frac{1}{2}\left(
\begin{array}{cccc}
i\nu -i\mu & \mu +\nu & it_{2}\left( 1-e^{-ik}\right) & t_{2}\left(
1+e^{-ik}\right) \\
\mu +\nu & i\mu -i\nu & t_{2}\left( 1+e^{-ik}\right) & -it_{2}\left(
1-e^{-ik}\right) \\
-it_{2}\left( 1-e^{ik}\right) & t_{2}\left( 1+e^{ik}\right) & i\mu -i\nu &
\mu +\nu \\
t_{2}\left( 1+e^{ik}\right) & it_{2}\left( 1-e^{ik}\right) & \mu +\nu & i\nu
-i\mu%
\end{array}%
\right) .
\end{equation}%
The Bloch Hamiltonians $H_{a}\left( k\right) $ and $H_{c}\left( k\right) $
are connected via $U_{ac}=\left( i\sigma _{x}+I_{2}\right) $ as%
\begin{equation}
U_{ac}H_{a}\left( k\right) U_{ac}^{-1}=H_{c}\left( k\right) ;
\end{equation}%
the Bloch Hamiltonians $H_{b}\left( k\right) $ and $H_{d}\left( k\right) $
are connected via $U_{bd}=I_{2}\otimes \left( i\sigma _{x}+I_{2}\right) $ as%
\begin{equation}
U_{bd}H_{b}\left( k\right) U_{bd}^{-1}=H_{d}\left( k\right) .
\end{equation}

For the sake of comparison, the Bloch Hamiltonians of systems $a$ and $c$
are alternatively shown in form of
\begin{equation}
H_{a}^{\prime }\left( k\right) =\left(
\begin{array}{cccc}
0 & \mu & 0 & t_{2}e^{-ik} \\
\nu & 0 & t_{2} & 0 \\
0 & t_{2} & 0 & \mu \\
t_{2}e^{ik} & 0 & \nu & 0%
\end{array}%
\right) ,H_{c}^{\prime }\left( k\right) =\frac{1}{2}\left(
\begin{array}{cccc}
i\nu -i\mu & \mu +\nu & it_{2}\left( 1-e^{-ik}\right) & t_{2}\left(
1+e^{-ik}\right) \\
\mu +\nu & i\mu -i\nu & t_{2}\left( 1+e^{-ik}\right) & -it_{2}\left(
1-e^{-ik}\right) \\
-it_{2}\left( 1-e^{ik}\right) & t_{2}\left( 1+e^{ik}\right) & i\nu -i\mu &
\mu +\nu \\
t_{2}\left( 1+e^{ik}\right) & it_{2}\left( 1-e^{ik}\right) & \mu +\nu & i\mu
-i\nu%
\end{array}%
\right) ,
\end{equation}%
where two unit cells are considered as a compound one, $U_{bd}H_{a}^{\prime
}\left( k\right) U_{bd}^{-1}=H_{c}^{\prime }\left( k\right) $. In the
discrete systems with lattice size $N=4n$, the wave vector $k$ is $k=\pi m/n$%
, $m\in \lbrack 1,2n]$ ($m,n$ are positive integers) for the Bloch
Hamiltonians $H_{a}\left( k\right) $ and $H_{c}\left( k\right) $ with a
two-site unit cell, and the wave vector $k$ is $k=2\pi m/n$, $m\in \lbrack
1,n]$ for the Bloch Hamiltonians $H_{a}^{\prime }\left( k\right) $, $%
H_{b}\left( k\right) $, $H_{c}^{\prime }\left( k\right) $, and $H_{d}\left(
k\right) $ with a four-site unit cell.\ Notably, $H_{a}\left( k\right) $ and
$H_{a}^{\prime }\left( k\right) $ [$H_{c}\left( k\right) $ and $%
H_{c}^{\prime }\left( k\right) $] yield identical energy bands.

\subsection*{I. 2D non-Hermitian topological systems}

The breakdown and recovery of conventional bulk-boundary correspondence are
discussed for two 2D non-Hermitian topological systems. The Bloch
Hamiltonian of a 2D non-Hermitian Chern insulator is given by
\begin{equation}
H_{a,CI}\left( k_{x},k_{y}\right) =\left( m+t\cos k_{x}+t\cos k_{y}\right)
\sigma _{x}+\left( t\sin k_{x}+i\gamma \right) \sigma _{y}+\left( t\sin
k_{y}\right) \sigma _{z}.  \label{CI}
\end{equation}%
Set $\mu =m+t\cos k_{y}+\gamma $, $\nu =m+t\cos k_{y}-\gamma $, we obtain%
\begin{equation}
H_{a,CI}\left( k_{x},k_{y}\right) =\left(
\begin{array}{cc}
t\sin k_{y} & \mu +te^{-ik_{x}} \\
\nu +te^{ik_{x}} & -t\sin k_{y}%
\end{array}%
\right) .
\end{equation}%
The coupling $\mu $-$\nu $ is asymmetric for $\gamma \neq 0$. The energy
bands are $E_{a,CI,\pm }=\pm \sqrt{\mu \nu +t^{2}+t^{2}\sin
^{2}k_{y}+t\left( \mu e^{ik_{x}}+\nu e^{-ik_{x}}\right) }$, which can be
rewritten as
\begin{equation}
E_{a,CI,\pm }\left( k_{x},k_{y}\right) =\pm \sqrt{\mu \nu +t^{2}+t^{2}\sin
^{2}k_{y}+2t\sqrt{\mu \nu }\cos \left( k_{x}+i\phi \right) },  \label{ECI}
\end{equation}%
where $\sqrt{\mu /\nu }=e^{-\phi }$ for $\mu \nu >0$ [in the case of $\mu <0$
and $\nu >0$, we set $e^{-\phi }=i\sqrt{\left\vert \mu /\nu \right\vert }$;
in the case of $\mu >0$ and $\nu <0$, we set $e^{-\phi }=-i\sqrt{\left\vert
\mu /\nu \right\vert }$. In both cases, we obtain the energy bands by
replacing $\sqrt{\mu \nu }$ with $i\sqrt{\left\vert \mu \nu \right\vert }$
in Eq.~(\ref{ECI})]. From the energy bands, we notice that a nonzero
imaginary magnetic flux exists in the $x$ direction; thus, the bulk-boundary
correspondence fails in the $x$ direction~\cite{KawabataCI}. A
chiral-inversion symmetry in the $y$ direction can be defined for the
non-Hermitian Chern insulator $a$, which does not prevent the nonzero
imaginary magnetic flux in the $x$ direction.

To recover conventional
bulk-boundary correspondence, we can enforce an inversion symmetry when
introducing the non-Hermiticity (asymmetric coupling) in the Chern
insulator. Then, the bulk Bloch Hamiltonian is given by
\begin{equation}
H_{b,CI}\left( k_{x},k_{y}\right) =\left(
\begin{array}{cccc}
t\sin k_{y} & \sqrt{\mu \nu }e^{-\phi } & 0 & te^{-ik_{x}} \\
\sqrt{\mu \nu }e^{\phi } & -t\sin k_{y} & t & 0 \\
0 & t & t\sin k_{y} & \sqrt{\mu \nu }e^{\phi } \\
te^{ik_{x}} & 0 & \sqrt{\mu \nu }e^{-\phi } & -t\sin k_{y}%
\end{array}%
\right) .
\end{equation}%
The Bloch Hamiltonian $H_{b,CI}\left( k_{x},k_{y}\right) $ has zero
imaginary magnetic flux under PBC because of the cancellation between
amplification and attenuation factors $e^{\pm \phi }$ in the $x$ direction.
A chiral-inversion symmetry in the $x$ direction can be defined for the
non-Hermitian Chern insulator $b$. After a gauge transformation $U_{CI}=%
\mathrm{diag}\left( e^{ik_{x}/2},e^{ik_{x}/2},1,1\right) $, we obtain $%
H_{b,CI}^{\prime }\left( k_{x},k_{y}\right) =U_{CI}H_{b,CI}\left(
k_{x},k_{y}\right) U_{CI}^{-1}$,\ then $\left( \mathcal{SP}\right)
H_{b,CI}^{\prime }\left( k_{x},k_{y}\right) \left( \mathcal{SP}\right)
^{-1}=-H_{b,CI}^{\prime }\left( -k_{x},k_{y}\right) $ is satisfied with $%
\mathcal{SP}=\sigma _{y}\otimes \sigma _{y}$. By applying the procedure done
for the SSH model with asymmetric coupling, we can obtain an equivalent bulk
Bloch Hamiltonian through removing the imaginary gauge field (wiping off the
amplification and attenuation factors $e^{\pm \phi }$), which gives%
\begin{equation}
h_{b,CI}\left( k_{x},k_{y}\right) =\left(
\begin{array}{cc}
t\sin k_{y} & \sqrt{\mu \nu }+te^{-ik_{x}} \\
\sqrt{\mu \nu }+te^{ik_{x}} & -t\sin k_{y}%
\end{array}%
\right) .
\end{equation}%
The energy bands are
\begin{equation}
E_{b,CI,\pm }\left( k_{x},k_{y}\right) =\pm \sqrt{\mu \nu +t^{2}+t^{2}\sin
^{2}k_{y}+2t\sqrt{\mu \nu }\cos \left( k_{x}\right) }.
\end{equation}%
The bulk topology of $h_{b,CI}\left( k_{x},k_{y}\right) $ can correctly
predict the topological phase transition and the (non)existence of edge
states for both Chern insulators $a$ and $b$ under OBC.

The approach is applicable in a 2D Rice-Mele model studied in Ref.~\cite%
{Kunst}. The Bloch Hamiltonian is given by
\begin{equation}
H_{a,RM}\left( k_{x},k_{y}\right) =\left[ t_{1}+\delta \cos k_{x}+\left(
t_{1}-\delta \cos k_{x}\right) \cos k_{y}\right] \sigma _{x}+\left[ \left(
t_{1}-\delta \cos k_{x}\right) \sin k_{y}+i\gamma /2\right] \sigma
_{y}-\left( \Delta \sin k_{x}\right) \sigma _{z}.
\end{equation}%
Set $\mu =t_{1}+\delta \cos k_{x}+\gamma /2$, $\nu =t_{1}+\delta \cos
k_{x}-\gamma /2$, and $\sqrt{\mu /\nu }=e^{-\phi }$ for $\mu \nu >0$, we
obtain
\begin{equation}
E_{a,RM,\pm }=\pm \sqrt{\mu \nu +\left( t_{1}-\delta \cos k_{x}\right)
^{2}+\Delta ^{2}\sin ^{2}k_{x}+2\left( t_{1}-\delta \cos k_{x}\right) \sqrt{%
\mu \nu }\cos \left( k_{y}+i\phi \right) }.
\end{equation}%
In contrast to the non-Hermitian Chern insulator shown in Eq. (\ref{CI}), a
nonzero imaginary magnetic flux exists in the $y$ direction, but not in the $%
x$ direction of the 2D Rice-Mele model. Thus, in the $y$ direction, the PBC
and OBC spectra considerably differ from each other and conventional
bulk-boundary correspondence fails in the $y$ direction~\cite{Kunst}. By
enforcing an inversion symmetry when introducing the non-Hermiticity
(asymmetric coupling) in the system, the imaginary magnetic flux vanishes,
the conventional bulk-boundary correspondence recovers in the $y$ direction.
The bulk Bloch Hamiltonian has zero imaginary magnetic flux and is given by%
\begin{equation}
H_{b,RM}\left( k_{x},k_{y}\right) =\left(
\begin{array}{cccc}
-\Delta \sin k_{x} & \sqrt{\mu \nu }e^{-\phi } & 0 & \left( t_{1}-\delta
\cos k_{x}\right) e^{-ik_{y}} \\
\sqrt{\mu \nu }e^{\phi } & \Delta \sin k_{x} & \left( t_{1}-\delta \cos
k_{x}\right) & 0 \\
0 & \left( t_{1}-\delta \cos k_{x}\right) & -\Delta \sin k_{x} & \sqrt{\mu
\nu }e^{\phi } \\
\left( t_{1}-\delta \cos k_{x}\right) e^{ik_{y}} & 0 & \sqrt{\mu \nu }%
e^{-\phi } & \Delta \sin k_{x}%
\end{array}%
\right) .
\end{equation}%
The equivalent system is obtained through removing the imaginary gauge
field. Then, we obtain
\begin{equation}
h_{b,RM}\left( k_{x},k_{y}\right) =\left(
\begin{array}{cc}
-\Delta \sin k_{x} & \sqrt{\mu \nu }+\left( t_{1}-\delta \cos k_{x}\right)
e^{-ik_{y}} \\
\sqrt{\mu \nu }+\left( t_{1}-\delta \cos k_{x}\right) e^{ik_{y}} & \Delta
\sin k_{x}%
\end{array}%
\right) .
\end{equation}%
The energy bands are given by
\begin{equation}
E_{b,RM,\pm }=\pm \sqrt{\mu \nu +\left( t_{1}-\delta \cos k_{x}\right)
^{2}+\Delta ^{2}\sin ^{2}k_{x}+2\left( t_{1}-\delta \cos k_{x}\right) \sqrt{%
\mu \nu }\cos \left( k_{y}\right) }.
\end{equation}%
The bulk topology of $h_{b,RM}\left( k_{x},k_{y}\right) $ can correctly
predict the topological phase transition and the (non)existence of edge
states for both Rice-Mele models $a$ and $b$ under OBC.

\clearpage
\end{widetext}

\begin{thebibliography}{999}
\bibitem{RemarkPeriels} Peierls phase factor refers $e^{\pm i\Phi }$~\cite%
{Goldman}, where $\Phi $ is imaginary $\Phi =i\phi $ for the asymmetric
coupling $\sqrt{\mu \nu }e^{\mp \phi }$.
\bibitem{ChiralInversionSymmetry} For system $b$, $\mathcal{SP=}\sigma
_{x}\otimes \sigma _{y}$ and $U_{\mathcal{SP}}=P_{2n}\otimes \sigma _{y}$ in
Eq.~(\ref{2}), where $\otimes $ is the Kronecker product, and $P_{2n}$ is a $%
90$ degrees rotation of the $2n\times 2n$ identical matrix $I_{2n}$.
\bibitem{NoteBoundary} If the asymmetric intracell coupling $\mu $-$\nu $ at
system boundary vanishes instead of the intercell coupling $t_{2}$, the
conclusion for the existence of zero edge states is inverse as in the
regions $t_{1}^{2}<\gamma ^{2}-t_{2}^{2}$ and $t_{1}^{2}>\gamma
^{2}+t_{2}^{2}$.
\end{thebibliography}
\end{document}